\documentclass[10pt,twocolumn,pra,aps,floatfix,superscriptaddress,longbibliography]{revtex4-1}

\usepackage{color}
\usepackage{graphicx}
\usepackage{physics}
\usepackage{amsthm}
\usepackage{amsmath}
\usepackage{amssymb}
\usepackage{enumerate}
\usepackage{placeins}
\usepackage{booktabs}
\usepackage{dsfont}
\usepackage{yfonts}
\usepackage{hyperref}

\newcommand{\Eqref}[1]{Eq.~\eqref{#1}}

\usepackage[color=green!60]{todonotes}

\AtBeginDocument{%
\heavyrulewidth=.08em
\lightrulewidth=.05em
\cmidrulewidth=.03em
\belowrulesep=.65ex
\belowbottomsep=0pt
\aboverulesep=.4ex
\abovetopsep=0pt
\cmidrulesep=\doublerulesep
\cmidrulekern=.5em
\defaultaddspace=.5em
}

\begin{document}

\title{Equivalence of Spatial and Particle Entanglement Growth After a Quantum Quench}

\author{Adrian Del Maestro}
\affiliation{Department of Physics and Astronomy, University of Tennessee, Knoxville, TN 37996, USA}
\affiliation{Min H. Kao Department of Electrical Engineering and Computer Science, University of Tennessee, Knoxville, TN 37996, USA}

\author{Hatem Barghathi}
\affiliation{Department of Physics and Astronomy, University of Tennessee, Knoxville, TN 37996, USA}

\author{Bernd Rosenow}
\affiliation{Institut f\"ur Theoretische Physik, Universit\"at Leipzig, D-04103, Leipzig, Germany} 

\begin{abstract}
We analyze fermions after an interaction quantum quench in one spatial dimension and study the growth of the steady state entanglement entropy density under either a spatial mode or particle bipartition. For integrable  lattice models, we find excellent agreement between the increase of spatial and particle entanglement entropy, and for chaotic models, an examination of two further neighbor interaction strengths suggests similar correspondence.  This result highlights the generality of the dynamical conversion of entanglement to thermodynamic entropy under time evolution that underlies our current framework of quantum statistical mechanics.    
\end{abstract}

\maketitle

\section{Introduction}
\label{sec:introduction}

The time evolution of an initial  quantum state after a sudden change of interaction strength  leads to an asymptotic steady state, whose local properties are governed by the buildup of entanglement between spatial subregions of the system \cite{Srednicki:1994xg, Rigol:2008sl, Calabrese:2005oi, Polkovnikov:2011gg, DAlessio:2016rr, Alba:2017ph}.  This entanglement is believed to be responsible for the generation of extensive entropy that validates the use of statistical mechanics for local expectation values,  an idea which is  supported by recent measurements of the 2nd R\'{e}nyi entropy \cite{Kaufman:2016ep,Lukin:2018wg}.  The ability to experimentally investigate the unitary time evolution of pure states in isolated systems on long time scales in ultra-cold atoms \cite{Kinoshita:2006yz,Langen:2013ls,Kaufman:2016ep} now provides an exciting opportunity to test fundamental ideas on how quantum statistical mechanics emerges from the many-body time dependent Schr\"odinger equation. 

\begin{figure}[!t]
\includegraphics[width=0.82\columnwidth]{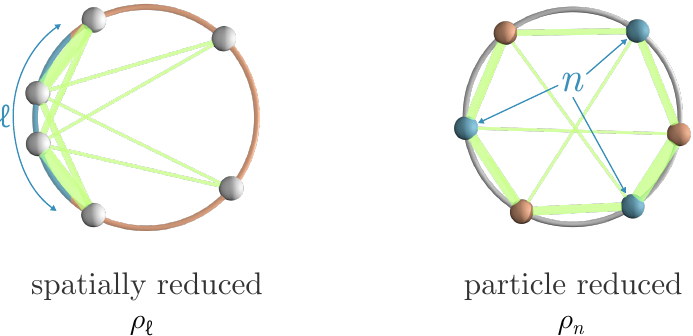}
\caption{\label{fig:bipartitions}%
Two types of reduced density matrices after a quantum quench.
A quantum system of interacting fermions in one-dimension with periodic boundary conditions can be bipartitioned into 
 a spatial partition of size $\ell$ (left) or a particle partition consisting of $n$ fermions (right).  The degrees of freedom which are kept in the reduced density matrix are indicated in blue, while the orange ones are traced out. Interactions between the former and latter are pictured with green lines.}
\end{figure}

As an alternative to the conventional spatial mode partitioning, a quantum system of $N$ indistinguishable particles can be bipartitioned into two groups containing $n$ and $N-n$  particles each \cite{Haque:2007il,Zozulya:2007jw,Zozulya:2008kb, Haque:2009df, Liu:2010pe, Herdman:2014jq, Herdman:2014ey, Herdman:2015gx, Rammelmuller:2017om, Barghathi:2017fv,Hopjan:2020sp} as shown in Fig.~\ref{fig:bipartitions}. The $n$-particle reduced density matrix $\rho_n$ can be computed in practice by keeping $n$ particle coordinates fixed while tracing over the remaining $N-n$ particle positions in the appropriately symmetrized first-quantized wavefunction \cite{Lowdin:1955uo,Coleman:1963lt, Mahan:2000mp}.  In this way, the partial trace is performed while fully respecting the indistinguishably of quantum particles.  The elements of this reduced density matrix are proportional to correlation functions, and are thus in principle measurable in experiments, and the resulting entanglement entropy has been shown to be sensitive to both interactions and particle statistics at leading order \cite{Zozulya:2008kb, Herdman:2015gx,Barghathi:2017fv}.

A general finite size scaling form has been conjectured for the ground state particle entanglement (von Neumann entropy of $\rho_n$) of interacting systems \cite{Zozulya:2008kb, Haque:2009df, Barghathi:2017fv} that behaves like $n \ln N$ ($n \ll N$),  
markedly  different from the area law $\ell^{D-1}$ of spatial entanglement for subregion size $\ell$ in dimension $D$ for gapped quantum systems with reduced density matrix $\rho_\ell$ with local interactions \cite{Hastings:2007bu,Eisert:2010hq}. While there has been renewed interest in entanglement dynamics for non-spatial single-particle bipartitions \cite{Pastori:2019ds,Lundgren:2019aa}, little is known about the evolution of entanglement between groups of particles after a quantum quench.

%
%
%
 In this paper, we compare the growth of the steady-state entanglement entropy  after a quantum quench under spatial and particle bipartations, for both integrable and chaotic models of one-dimensional interacting lattice fermions. By fully exploiting translational symmetries of particle subgroups,  we exactly determine large 6-particle reduced density matrices for systems containing up to $L=26$ sites at half-filling, making a well-controlled extrapolation to the thermodynamic limit possible. Having access to the thermodynamic limit via finite size scaling, we find agreement between the asymptotic increase  of entropy densities computed from spatial and particle bipartitions. This equivalence with respect to the specific partition of the quantum state supports the notion that  the properties  of a steady state local equilibrium are fundamental to a statistical mechanics description of many particle systems. 

The paper is organized as follows: after introducing the details of our model and quantum quench protocol in Section~II, we discuss the definition of spatial and particle entanglement dynamics in Section~III and the approach to extracting the asymptotic entanglement density in Section~IV.  Our main results for both integrable and non-integrable models are contained in Section V, and we conclude by discussing implications our results in Section VI.

\section{Quantum Quench}
We study a system of $N$ spinless fermions on $L$ lattice sites in one spatial dimension (1D) 
with hopping amplitude $J$ and time-dependent nearest,  $V(t)$,  and next-nearest neighbor $V^\prime(t)$ interactions described by the Hamiltonian 
\begin{align}
    H &= -J\sum_{i=1}^{L}\left(c^\dagger_{i} c^{\phantom{\dagger}}_{i+1} +c^\dagger_{i+1} 
    c^{\phantom{\dagger}}_{i} \right) + V(t)\sum_{i=1}^{L} n_i n_{i+1}  \nonumber \\
      &\quad + V^\prime(t)\sum_{i=1}^{L} n_i n_{i+2}
  \label{eq:H-JV}
\end{align}
where $c^\dagger_{i}(c^{\phantom{\dagger}}_i)$ creates(annihilates) a fermion on site $i$, $\{c_i^{\phantom{\dagger}},c_j^\dagger\} = \delta_{ij}$ and $n_i = c_i^\dagger c_i$ is the occupancy of site $i$.  
For $V'(t) = 0$, Eq.~(\ref{eq:H-JV}) can be mapped onto the XXZ spin-$\tfrac{1}{2}$ chain at fixed total spin $S$ which is exactly solvable via Bethe ansatz \cite{DesCloizeaux:1966,Yang:1966yc}.  In what follows we will measure all energies in units of the hopping $J$.

The system is prepared at times $t<0$ in an initial state of non-interacting spinless fermions with 
\begin{equation}
\ket{\Psi(0)} = \prod_{k \leq k_{\rm F}} c_k^\dagger \ket{0} 
\label{eq:Psi0}
\end{equation}
where $\ket{0}$ is the vacuum state.  We employ periodic boundary conditions for odd $N$ and anti-periodic for even $N$ to avoid complications arising from a possibly degenerate ground state such that the lattice Fourier transform picks up a phase in the anti-periodic case:
\begin{align}
    c^{\phantom \dagger}_k &=  \frac{1}{\sqrt{L}}\sum_{j=1}^{L} c^{\phantom \dagger}_j \mathrm{e}^{-\imath k j}
    \begin{cases}
        1 & N\ \text{odd} \\
        \mathrm{e}^{\imath \pi j/L} & N\ \text{even} 
    \end{cases}\,.
\label{eq:LatticeFourierTransform}
\end{align}
The quasi-momenta are 
\begin{align}
    k \in 
    \begin{cases}
        -\pi(N-1)/L,\dots, \pi(N-1)/L & N\ \text{odd} \\
        -\pi N/L+2\pi/L,\dots,\pi N/L & N\ \text{even}
    \end{cases}
\end{align}
such that the Fermi momentum is $k_{\rm F} = \pi \tfrac{N}{L} - \pi\tfrac{1-(-1)^N}{2L}$.

At time $t=0$, interactions of strength $V$ and $V^\prime$ are turned on ($V(t) = V \Theta(t), V^\prime(t) = V^\prime\, \Theta(t)$ with $\Theta$ the Heaviside step function). 
The state of the system at time $t$ after the quench is given by unitary time evolution of $\ket{\Psi(0)}$ under $H$:
\begin{equation}
    \ket{\Psi(t)} = \mathrm{e}^{-\imath H t} \ket{\Psi(0)} =  \sum_\alpha \mathrm{e}^{-\imath E_\alpha t} \braket{\Psi_\alpha}{\Psi(0)} \ket{\Psi_\alpha}
\label{eq:Psit}
\end{equation}
where we have set $\hbar=1$ and $E_\alpha$ and $\ket{\Psi_\alpha}$ are the energy eigenvalues and eigenstates of the post-quench Hamiltonian, $H \ket{\Psi_\alpha} = E_\alpha \ket{\Psi_\alpha}$, obtained from full exact diagonalization exploiting the translational, inversion, and particle-hole symmetry of \Eqref{eq:H-JV}. All software, data, and scripts needed to reproduce the results in this paper are available online \cite{repo}.  

\section{Entanglement Dynamics}

Tracing out spatial degrees of freedom outside of a contiguous region of $\ell$ sites from the time-dependent density matrix
$\rho(t) = |\Psi(t)\rangle \langle \Psi(t)|$ yields the spatially reduced $\rho_\ell(t) = \Tr_{L-\ell} \rho(t)$.  For a particle bipartition,  the reduced density matrix $\rho_n(t)$ can be computed by fixing $n$ coordinates in the properly symmetrized many-particle wavefunction $\Psi(i_1,\dots,i_N;t) = \braket{i_1,\dots,i_N}{\Psi(t)}$ and tracing over the remaining $N-n$ positions: 
%
\begin{multline}
\!\!\!\!\! \rho_n^{i_1,\dots,i_n;j_1,\dots,j_{n}}(t) = \!\!
    \sum_{i_{n+1},\dots,i_N}\!\! \Psi^\ast\qty(i_1,\dots,i_{n},i_{n+1},\dots,i_{N};t)
    \\
    \times \Psi\qty(j_1,\dots,j_{n},i_{n+1},\dots,i_{N};t)
   \label{eq:nrdm}
\end{multline}
where the particle coordinates $i_1 \dots,i_N$ can take any position on the lattice. 
A graphical comparison of their entanglement structure in real space is depicted in Fig.~\ref{fig:bipartitions}.  

The von Neumann entanglement entropy at each time $t$ is computed from 
the spatial ($\ell$) or particle ($n$) reduced density matrix
\begin{equation}
    S(t;n\vert\ell) = -{\rm Tr} \left[ \rho_{n\vert\ell}(t) \ln \rho_{n\vert\ell}(t)\right]\,. 
\label{eq:renyi}
\end{equation}
In gapless 1D quantum systems after a global quantum quench,  the entanglement entropy under a spatial bipartition of length $\ell$ grows linearly with time $t$: $S \propto t$ up to $t = \ell/(2v)$, and then saturates to a value that is extensively large in the sub-region size: $S \propto \ell /(2v )$ \cite{Calabrese:2005oi,Fioretto:2010qc,Stephan:2011lq} were $v$ is a velocity. This can be understood in terms of the stimulated emission of highly entangled quasi-particles inside the sub-region that propagate outwards with $v$. Many of these results have been tested against numerical calculations on
lattice models starting from unentangled product states
\cite{Lauchli:2008wb,Calabrese:2016sn,Alba:2017ph,Garrison:2018kv} highlighting the regime of applicability of conformal field theory. 

For particle entanglement, the reduced density matrix $\rho_n$ has $L^{2n}$ elements (see Eq.~(\ref{eq:nrdm})), but due to the indistinguishability of particles, the effective linear size of the matrix size is only $\binom{L}{n}$ \cite{Barghathi:2017fv}. Even with this reduction, for $N=13$ and $n=6$ at half-filling ($L=26$), the determination of its dynamics requires the full diagonalization of a matrix with over $5\times 10^{10}$ elements at each time step.  This would make an exact analysis of the steady state particle entanglement in the thermodynamic limit computationally intractable.  Only when reducing the number of matrix elements  additional factor of $\sim L^2$ by exploiting translational invariance within particle subgroups (see Appendix \ref{app:ParticleEntanglement}), does the numerical diagonalization becomes feasible.  Our software level implementation of this approach for the $n$-particle reduced density matrices considered here is included in an online repository~\cite{repo}. 

\section{Entanglement Density}
As we are interested in the initial growth, and final asymptotic steady state value of entanglement entropy, we study the difference between its value at an observation time $t$ after the quench, and that of the initial pre-quench non-interacting fermionic state: 
\begin{equation}
\Delta S(t;n\vert\ell) \equiv S(t;n\vert\ell) - S(0;n\vert\ell).   
\label{eq:DeltaS}
\end{equation}
This removes the $t=0$ contribution of the free fermion state described by a single Slater determinant \cite{Jin:2004uy,Ghirardi:2005md,Carlen:2016es,Lemm:2017ef}: 
\begin{equation}
\begin{aligned}
    S(0;\ell) &= \frac{1}{3}\ln \frac{2\ell}{\pi} + 0.495\dots  \\
    S(0;n) &= \ln \binom{N}{n} \, .
\end{aligned}
\label{eq:freeFermionEntanglement}
\end{equation}
It is useful to point out the different scaling properties of these two quantities in equilibrium.  While the spatial mode entanglement behaves as $S(0;\ell) \sim \ln \ell$, for $n\ll N$, $S(0;n) \sim n \ln N$, and moreover, while the former is insensitive to particle statistics, the particle entanglement arises purely from the anti-symmetrization of the wavefunction in first quantization (it is exactly zero for non-interacting bosons \cite{Haque:2009df,Herdman:2016ep}).   Due to this 
qualitative difference in the system size dependence of the initial state entanglement entropy, it is important to compare the {\em difference} between 
the asymptotic and initial state entanglement entropies with regards to spatial and particle bipartitions after the quench. 

\begin{figure}[t]
\begin{center}
\includegraphics[width=1.0\columnwidth]{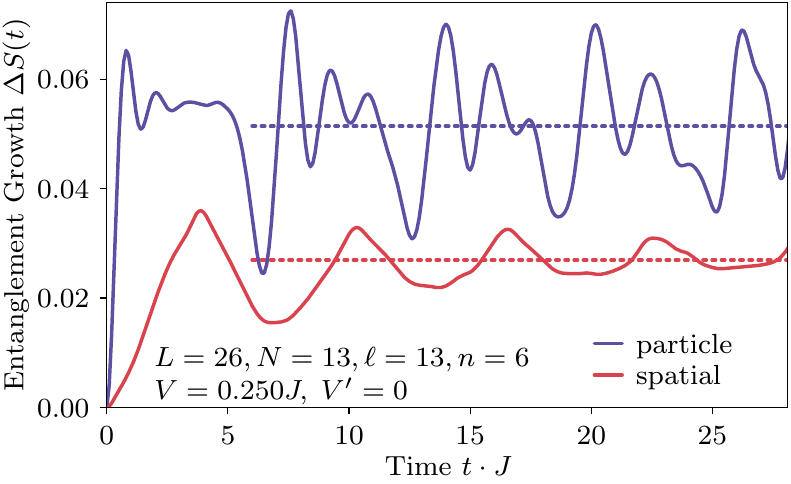}
\end{center}
\caption{{Exact diagonalization results for entanglement.} Time-dependence of 
    the increase in  particle and spatial entanglement entropy after an interaction of strength $V=0.25J$ is turned on ($V^\prime = 0$). The spatial entanglement entropy (red curve) for $\ell = L/2=13$ sites
grows linearly up to a time $t J \sim \ell/4$, whereas
the particle entanglement entropy (purple curve) for $n=\lfloor N/2\rfloor=6$ particles rapidly increases on a scale $t J \sim 1/2$. The dashed lines show the asymptotic $t\to\infty$ values extracted from the full time dependence (data included in Ref.~\cite{repo}).}
\label{fig:ED}
\end{figure}

In Fig.~\ref{fig:ED}, the time dependence of $\Delta S(t)$ for both spatial and particle bipartitions is shown for a system with $N=13$ particles on $L=26$ sites (half-filling), 
%
%
for maximal bipartition sizes of $n=6$ particles and $\ell=13$ sites with $V^\prime = 0$.  Even/odd parity effects in the particle entanglement entropy can be mitigated by replacing 
\begin{equation}
S(t;n) \to \frac{1}{2}\left[S(t;n) + \frac{n}{N-n}S(t;N-n)\right]
\label{eq:particle_normalization}
\end{equation}
when $n = \lfloor N/2 \rfloor$, where $\lfloor \dots \rfloor$ denotes the integer part.  We observe that the particle entanglement entropy rises to a value larger than the asymptotic one (indicated by the dashed line) within a microscopic time scale $tJ \sim 1/2$, whereas spatial entanglement entropy rises over a longer time $tJ  \sim \ell/4$. The finite size value of the entanglement entropy is larger for particle  than for spatial entanglement, and the amplitude of oscillations around the asymptotic average (dashed line) is larger for particle entanglement as well.

An estimate for the asymptotic $t\to\infty$ steady state entanglement entropy was obtained via time averaging:
\begin{equation}
    \Delta S (t\to\infty;n\vert \ell) \simeq \frac{1}{t_{f}-t_{i}}\int_{t_{i}}^{t_{f}} \!\!\dd{t} \left[S(t;n\vert \ell) - S(0;n\vert \ell)\right] \ . 
\label{eq:timeAverage}
\end{equation}
The average is started from $t_{i}J = N/2$, to correspond to the first recurrence time (see Fig.~\ref{fig:ED}), and the maximal time $t_{f}J = 100$ was chosen such that the statistical uncertainty in $\Delta S$ obtained by a binning analysis (allowing us to estimate error bars) was less than $3.5\%$.

Results in the thermodynamic limit ($n,\ell \to \infty$ such that $n/N,\ell/L \to \text{const.}$) can be obtained by fitting finite size exact diagonalization data for the maximal bipartition ($\ell = L/2, n = \lfloor N/2 \rfloor$) to the scaling ansatz:
\begin{equation}
\frac{1}{n}\Delta S\qty(t\to\infty) = \mathfrak{s} + \mathcal{C} \frac{\ln N}{N}\,
\label{eq:fssS}
\end{equation}
where $\mathfrak{s}$ is the desired entropy density, $n$ is the number of particles in the sub-region, and $\mathcal{C}$ is a constant.   This choice was motivated by an expansion of the equilibrium free fermion particle entanglement for $n=\lfloor N/2\rfloor$ for large $N$:
\begin{equation}
\frac{2}{N} \ln \binom{N}{N/2}
\simeq 2 \ln 2 - \frac{1}{N}\ln N + \mathrm{O}\qty(\frac{1}{N})\, ,
\end{equation}
further demonstrating the importance of subtracting off an extensive contribution originating from the pre-quench ground state. 

\section{Equivalence of Asymptotic Entanglement Growth}

\begin{figure}[t]
\begin{center}
\includegraphics[width=1.0\columnwidth]{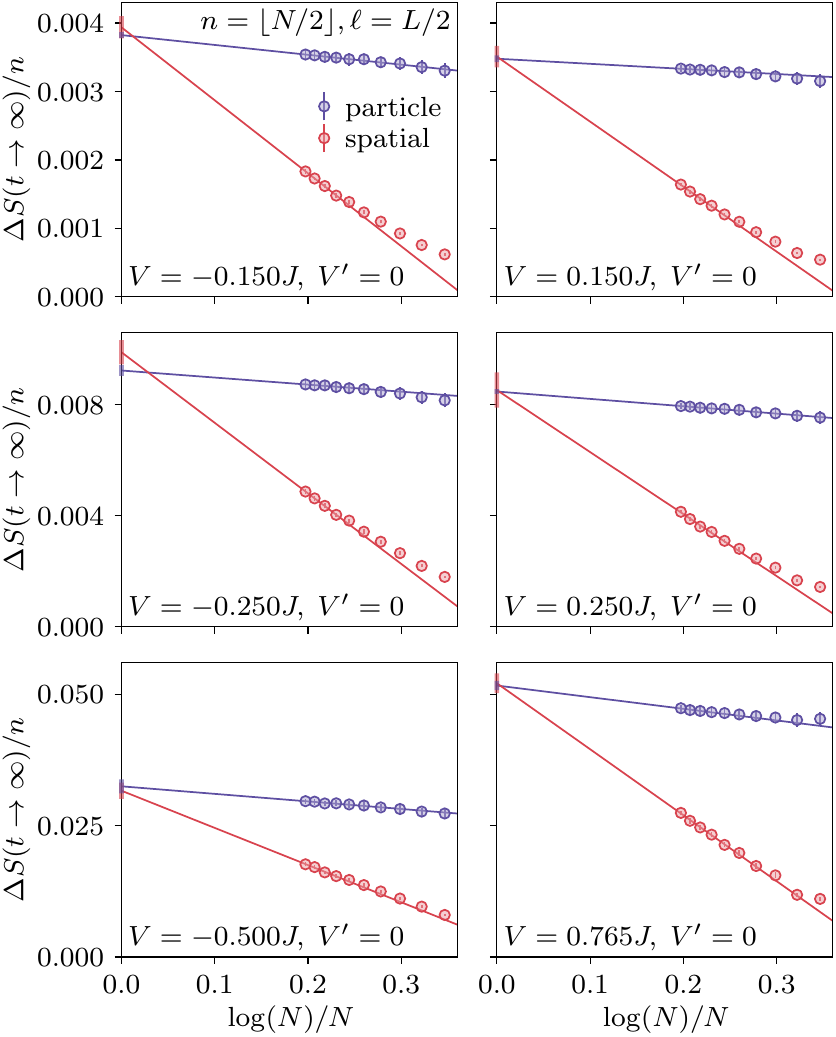}
\end{center}
\caption{{Asymptotic entanglement.} Finite size scaling of the $t\to\infty$ entanglement entropy per particle in the maximal subregion corresponding to $n=\lfloor N/2 \rfloor$ particles or $\ell = L/2$ sites for different nearest neighbor interactions $V$ and $V^\prime = 0$. Symbols correspond to exact diagonalization data and lines are fits to the finite size scaling form of Eq.~(\ref{eq:fssS}). Within the statistical uncertainty (size of shaded region) for $N\to\infty$, particle and spatial entanglement entropy density extrapolate to the same interaction dependent value $\mathfrak{s}$ in the thermodynamic limit.} 
\label{fig:universality}
\end{figure}

We begin by analyzing the quench of Eq.~\eqref{eq:H-JV} for the integrable case with $V^\prime = 0$.  We use Eq.~(\ref{eq:fssS}) to fit exact diagonalization data for both repulsive and attractive nearest neighbor interactions $V$, and thus obtain $\mathfrak{s}$. The uncertainty in $\mathfrak{s}$ is composed of two parts: the propagated error bars in the linear fit to the largest four system sizes ($N = 10,11,12,13$), and a possible systematic error due to the neglect of higher order terms in Eq.~\eqref{eq:fssS}.  The latter was estimated by computing the difference between the $N\to\infty$ extrapolated value for this fit, and two additional fits including $N=11,12,13$, or $N=9,10,11,12,13$ and averaging the resulting squared deviations.  The results of this combined finite size scaling and fitting procedure are shown in Fig.~\ref{fig:universality}, where $\mathfrak{s}$ corresponds to the line intercepts as $N\to\infty$. We find agreement within error bars between particle and spatial bipartitions in the thermodynamic limit.
%
%
Thus, we conclude that for the integrable case with $V^\prime = 0$, the asymptotic entanglement entropy per particle after an interaction quantum quench is equivalent under both a spatial and a particle bipartition in the thermodynamic limit.

\begin{figure}
\begin{center}
\includegraphics[width=1.0\columnwidth]{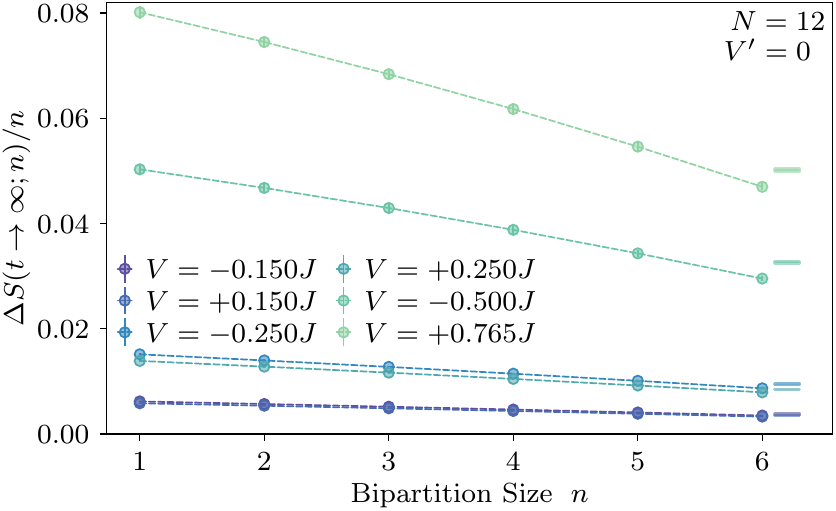}
\end{center}
\caption{{$n$-dependence of the particle entanglement entropy.} Bipartition size ($n$) dependence of the particle entanglement for $N=12$ particles on $L=24$ sites for various nearest neighbor interaction strengths $V$ with $V^\prime = 0$.  The $n$-particle entanglement entropy density is only weakly dependent on the order of the reduced density matrix.  Lines on the right hand side of the figure show the thermodynamic limit value of $\mathfrak{s}$ ($n,N\to\infty$ with $n/N = 1/2$) extracted from the fit shown in Fig.~\ref{fig:universality}.  Dashed lines are guides to the eye.}
\label{fig:nparticleEE}
\end{figure}

Interestingly,  we find that finite size corrections are much smaller  for  particle entanglement than for spatial entanglement \cite{Piroli:2017un}.  To explore this effect further, we keep $N$ fixed and study  how particle entanglement entropy  changes with $n$. The results are shown in Fig.~\ref{fig:nparticleEE} where $\tfrac{1}{n}\Delta S(t\to \infty;n)$ monotonically decreases with increasing $n$ \cite{Lemm:2017ef}.  
%
%
This effect can be explained by considering the growing number of constraints as correlations of up to $n$ particles are taken into account, with fewer states realizing the same reduced density matrix. Moreover, the existence of an non-monotonic entanglement shape function arising from $\rho_n = \rho_{N-n}$ implies a sub-linear growth of $\Delta S(n)$ for $n \approx N/2$, and thus a decrease of $\Delta S(n)/n$.  The result is that the $n$-particle entanglement entropy density can be estimated from knowledge of only the few lowest order density matrices.

To better understand the general applicability of the observed agreement between entanglement entropy growth under different bipartitions, we now lift the integrability constraint on the time-evolution of the initial state due to the existence of an infinite number of conservation laws. This is accomplished by including a next-nearest neighbor interaction $V^\prime$ that is quenched simultaneously with $V$ at $t=0$.

\begin{figure}[t]
\begin{center}
\includegraphics[width=1.0\columnwidth]{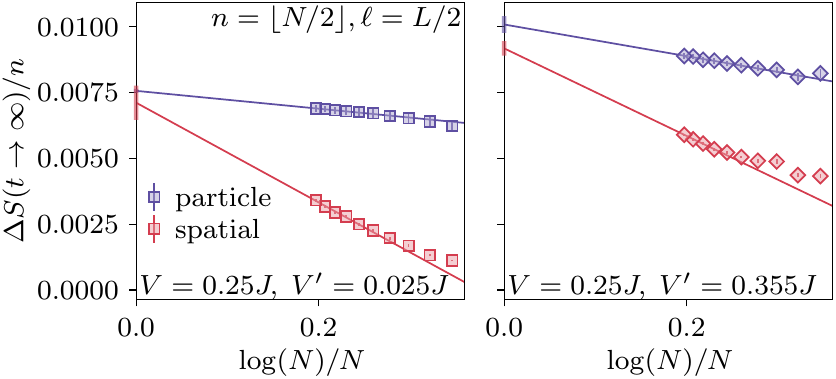}
\end{center}
\caption{Effects of integrability breaking. Finite size scaling of the spatial and particle entanglement entropy density for $t\to\infty$ for a fixed nearest neighbor interaction strength $V=0.25J$ with $V^\prime = 0.1 V$ (left) and $V^\prime = 1.42 V$ (right). Symbols correspond to exact diagonalization and lines are fits to Eq.~\eqref{eq:fssS}.} 
\label{fig:FSSVPrime}
\end{figure}

The equilibrium phase diagram of the $V$-$V^\prime $ model is known to be extremely complex \cite{Mishra:2011gd}, and we have chosen to fix $V = 0.25 J$ while investigating two next nearest neighbor interaction strengths $V^\prime = 0.025J$ and $V^\prime = 0.355 J$ to ensure we remain inside the quantum liquid phase and do not quench across a phase boundary.  Performing an  analysis identical to the integrable case above, we obtain the asymptotic post-quench finite size scaling results  shown in Fig.~\ref{fig:FSSVPrime}.
%
%
For weak $V^\prime/V = 0.1$ we find clear convergence to a common value of $\mathfrak{s}$ in the thermodynamic limit, while for extremely strong $V^\prime/V = 1.42$ equivalence is suggestive, but falls  outside the 1-$\sigma$ error bars.  
For both values of $V^\prime$, we find finite size effects to be more pronounced as compared to the integrable case (as expected due to the inclusion of a longer range interaction) and larger system sizes are required to enter the pure $\log N/N$ scaling regime.  Going to larger system sizes would be desirable, and while this is possible via the density matrix renormalization group \cite{White:1992rs,Schollwoeck:2011jf} for the spatial entanglement, at present, exact diagonalization remains the only viable route to obtain the spectra of $\rho_n$ for $n > 3$ \cite{Kurashige:2014ca}.

\begin{figure}[t]
\begin{center}
\includegraphics[width=1.0\columnwidth]{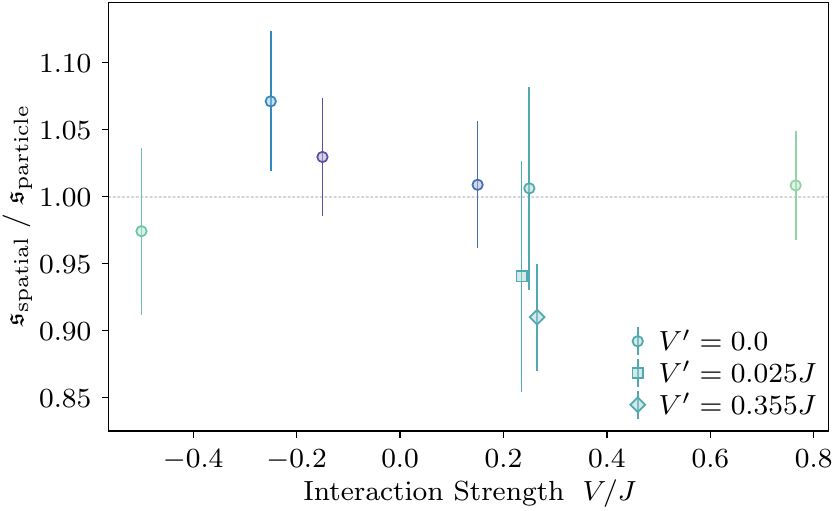}
\end{center}
\caption{Equivalence of spatial mode and particle entanglement. A comparison of the $t\to\infty$ and $N\to\infty$ entanglement density $\mathfrak{s}$ defined in Eq.~(\ref{eq:fssS}) (extrapolated values in Figs.~\ref{fig:universality} and \ref{fig:FSSVPrime}) as a function of quenched nearest-neighbor interaction strength $V$.  $V^\prime \ne 0$ points with $V=0.25J$ (square and diamond) have been horizontally shifted to better discern their error bars.}
\label{fig:extrapRelativeEE}
\end{figure}

Combining the extrapolated $t\to\infty$ and $N\to\infty$ results of Figs.~\ref{fig:universality} and \ref{fig:FSSVPrime} we can directly compare the prefactor $\mathfrak{s}$ of the extensive term in the asymptotic entanglement entropy under a spatial mode ($\mathfrak{s}_{\rm spatial}$) and particle ($\mathfrak{s}_{\rm particle}$) bipartition with the results shown in Fig.~\ref{fig:extrapRelativeEE}.
%
%
We observe agreement across a wide range of interactions spanning the entire quantum liquid regime including both attractive ($V<0$ and repulsive $V>0$) interactions, even in the presence of integrability breaking $V^\prime \ne 0$. We conclude that $\mathfrak{s}_{\rm spatial} \simeq \mathfrak{s}_{\rm particle}$ is consistent with the reported $\approx 5\%$ errorbars. For the non-integrable case with an extremely strong $V' > V$, agreement is within 10\%.  For this case, finite size effects are pronounced and exact diagonalization data may not yet be in the scaling regime causing an under reporting of uncertainty.  

\section{Discussion}

We have presented numerical results for an interaction quantum quench for both an integrable and non-integrable (chaotic) model of spinless fermions in one dimension, starting from a gapless and highly entangled non-interacting ground state.  Complementary to the often studied spatial entanglement entropy, we have examined a bipartition in terms of groups of particles, where the resulting entanglement can be obtained from the $n$-particle reduced density matrix.   At short times (as in equilibrium), the growth of entanglement  behaves very differently under these two bipartitions. In contrast, in the asymptotic long-time regime after subtracting the residual ground state value, we find an extensive entanglement entropy density that appears to be insensitive to the decomposition of the Hilbert space in terms of spatial or particle degrees of freedom. 
 
This equivalence  can be understood via the universal concept of coarse graining, a necessary ingredient of obtaining reduced density matrices that are  described by a generalized statistical ensemble  for integrable systems. While the computation of particle entanglement entropies after a quantum quench discussed in this study is not standard,  in other contexts the connection between particle reduced density matrices and entropy is well established. For example, in  equilibrium, the thermodynamic potential, and hence the entropy, can be computed from the one-particle density matrix when considering an adiabatic change of the coupling constant \cite{FeWa}. For classical non-equilibriums systems, according to Green \cite{Green52} and Kirkwood \cite{Kirkwood42} the distribution function can be factorized in an infinite hierarchy, enabling an expansion of the entropy in terms of irreducible correlation functions with increasing order. For classical liquids, it has been shown that even a termination of this entropy expansion at the pair level ($\rho_2$) is accurate to within 2\% \cite{Wallace87}, and for simulations of a system of soft disks this termination was shown to yield consistent results in non-equilibrium situations \cite{Evans89}.


\acknowledgments
We greatly benefited from discussions with F.~Heidrich-Meisner, A.~Polvkovnikov, and S.R.~White.  This work was supported in part by the NSF under Grant No.~DMR-1553991 and the DFG via grant RO 2247/11-1. H.B.\ acknowledges partial support from NSF Grant No.~DMR-1828489 during the manuscript editing phase. 

\appendix

\section{Translational Symmetry Resolved $n$-body Reduced Density Matrices}
\label{app:ParticleEntanglement}

In this appendix, we describe a specific example of how translational symmetry within particle subgroups can be exploited for $N=3$ fermions on a ring of $L=6$ sites and compute the spectrum of the resulting $1$-particle reduced density matrix. 

We begin by describing the decomposition of the occupation basis in terms of translational symmetry then discuss the Schmidt decomposition of a candidate ground state of Eq.~(\ref{eq:H-JV}) with $V=V'=0$ and finish with the explicit construction and resulting diagonalization of the reduced density matrix, comparing to its brute-force construction when no symmetries are taken into account.

\subsection{Translational Symmetry}
The translation operator $T$ acts as:
\begin{equation}
 T c_{i}^\dagger T^\dagger  =  c_{i+1}^\dagger,
\end{equation}
where, $c_{L+i}^\dagger =c_{i}^\dagger$, as we choose periodic boundary conditions for odd $N=3$.  Acting with $T$,  $L$ times will give back the same operator and thus $T^L$ is the identity. This immediately yields the eigenvalues of the unitary operator $T$ as $\mathrm{e}^{-\imath 2\pi q/L}$, where $q= 0,1,\dots,L-1$.

Now, consider the action of $T$ on the $L=6,N=3$ fermionic site-occupation basis $\qty{\ket{111000},\ket{011100},\dots}$ consisting of $\tbinom{L}{N} = 20$ states. Depending on how the indistinguishable particles are situated, the states can be grouped into $N_T = 4$ different types of translational cycles, each containing $M_\nu$ states for $\nu = 1\dots N_T$. 
The first 3 cycles all have $M_1 = M_2 = M_3 = 6$:
\begin{widetext}
\begin{align}
\ket{\psi_{1,1}} &= T^{0}\ket{\psi_{1,1}} =\ket{111000} & \ket{\psi_{2,1}} &= T^{0}\ket{\psi_{2,1}} =\ket{110100}&  \ket{\psi_{3,1}} &= T^{0}\ket{\psi_{3,1}} =\ket{110010}\nonumber\\
\ket{\psi_{1,2}} &= T^{1}\ket{\psi_{1,1}} =\ket{011100} & \ket{\psi_{2,2}} &= T^{1}\ket{\psi_{2,1}} =\ket{011010}&  \ket{\psi_{3,2}} &= T^{1}\ket{\psi_{3,1}} =\ket{011001}\nonumber\\
\ket{\psi_{1,3}} &= T^{2}\ket{\psi_{1,1}} =\ket{001110} & \ket{\psi_{2,3}} &= T^{2}\ket{\psi_{2,1}} =\ket{001101}&  \ket{\psi_{3,3}} &= T^{2}\ket{\psi_{3,1}} =\ket{101100}\nonumber\\
\ket{\psi_{1,4}} &= T^{3}\ket{\psi_{1,1}} =\ket{000111} & \ket{\psi_{2,4}} &= T^{3}\ket{\psi_{2,1}} =\ket{100110}&  \ket{\psi_{3,4}} &= T^{3}\ket{\psi_{3,1}} =\ket{010110}\nonumber\\
\ket{\psi_{1,5}} &= T^{4}\ket{\psi_{1,1}} =\ket{100011} & \ket{\psi_{2,5}} &= T^{4}\ket{\psi_{2,1}} =\ket{010011}&  \ket{\psi_{3,5}} &= T^{4}\ket{\psi_{3,1}} =\ket{001011}\nonumber\\
\ket{\psi_{1,6}} &= T^{5}\ket{\psi_{1,1}} =\ket{110001} & \ket{\psi_{2,6}} &= T^{5}\ket{\psi_{2,1}} =\ket{101001}&  \ket{\psi_{3,6}} &= T^{5}\ket{\psi_{3,1}} =\ket{100101}
\label{eq:l6cycles}
\end{align}
\end{widetext}
while the last one has $M_4 = 2$:
\begin{align}
    \ket{\psi_{4,1}} &= T^{0}\ket{\psi_{4,1}} =T^{2} \ket{\psi_{4,1}}=T^{4}\ket{\psi_{4,1}}=\ket{101010}\nonumber \\
    \ket{\psi_{4,2}} &= T^{1}\ket{\psi_{4,1}} = T^{3}\ket{\psi_{4,1}}= T^{5}\ket{\psi_{4,1}}=\ket{010101}
\label{eq:l2cycles}
\end{align}
where we have introduced new states $\ket{\psi_{\nu,m}}$ with $\nu=1\dots N_T$ and $m = 1\dots M_\nu$.  The eigenstates of $T$ can then be written as 
\begin{equation}
    \ket{\phi_{\nu,q}} = \frac{1}{\sqrt{M_\nu}} \sum_{m=1}^{M_\nu}\mathrm{e}^{\imath \frac{2\pi q}{M_\nu}(m-1)}\ket{\psi_{\nu,m}}\,,
\label{eq:phiTstates}
\end{equation}
where the corresponding eigenvalues are $\mathrm{e}^{-\imath 2\pi q /M_\nu}$ with $q=0,1,\dots,M_\nu-1$.

\subsection{Free fermion ground state}%
\label{sub:ground_state_of_the_j_v_model}

Consider Eq.~(\ref{eq:H-JV}) at $t\le0$ which corresponds to free lattice fermions ($V=V^\prime=0$).  The Hamiltonian possesses translational symmetry ($[T,H]=0$)
and thus the non-degenerate ground state $|\Psi_{0}\rangle$ must also be an eigenstate of the operator $T$.  Using the occupation basis states $\ket{\psi_{\nu,m}}$ introduced above, all matrix elements $\mel{\psi_{{\nu}^\prime,{m}^\prime}}{H}{\psi_{\nu,m}}$ of $H$ are real, and thus any non-degenerate eigenstate of $H$ must have real coefficients (up to an overall phase factor). This is only possible if the ground state is an eigenstate of $T$ with a real eigenvalue, 
\emph{i.e.}, $T\ket{\Psi_{0}}=\pm\ket{\Psi_{0}}$.  For free fermions, $\ket{\Psi_0}$ has zero total quasi momentum and thus $T\ket{\Psi_{0}}=+\ket{\Psi_{0}}$. Therefore, we can write  
\begin{equation}
    \ket{\Psi_{0}}=\sum_{\nu=1}^{N_T} a_\nu\ket{\phi_{\nu,0}}   
\label{eq:PsiTbasis}
\end{equation}
where $\sum_{\nu} a_\nu^2=1$. To evaluate the coefficients $a_\nu$ for free fermions, we consider the action of $H$ with $V=V^\prime=0$ on the states $\ket{\phi_{\nu,0}}$: 
\begin{align*}
H|\phi_{1,0}\rangle &=-J\left(|\phi_{2,0}\rangle +|\phi_{3,0}\rangle \right) \\
H|\phi_{2,0}\rangle &=-J\left(|\phi_{1,0}\rangle+2|\phi_{3,0}\rangle +\sqrt{3}|\phi_{4,0}\rangle \right) \\
H|\phi_{3,0}\rangle &=-J\left(|\phi_{1,0}\rangle+2|\phi_{2,0}\rangle +\sqrt{3}|\phi_{4,0}\rangle \right) \\
H|\phi_{4,0}\rangle &=-\sqrt{3}J\left(|\phi_{2,0}\rangle+|\phi_{3,0}\rangle \right)\,.
\label{eq:HPhi}
\end{align*}
Diagonalizing $H$ in this basis we find the ground state:
\begin{equation}
    \ket{\Psi_0} = \frac{\sqrt{3}}{6} \ket{\phi_{1,0}} +  \frac{\sqrt{3}}{3} \qty(\ket{\phi_{2,0}} + \ket{\phi_{2,0}}) + \frac{1}{2}\ket{\phi_{4,0}}
\label{eq:GST}
\end{equation}
and thus identify $a_1=\frac{\sqrt{3}}{6}$, $a_2=a_3=\frac{\sqrt{3}}{3}$ and $a_4=\frac{1}{2}$. As the introduction of the interaction terms in the Hamiltonian post-quench does not break translational symmetry, we are guaranteed to remain in the $q=0$ sector and thus the general state $\ket{\Psi(t)}$ can always be decomposed as in Eq.~(\ref{eq:PsiTbasis}), however the time-dependent coefficients may now be complex in general. 

\subsection{Schmidt decomposition of the ground state}%
\label{ssub:schmidt_decomposition_of_the_ground_state}

In order to perform a particle bipartition, we first need to artificially distinguish the identical fermions from each other, i.e., we write the ground state in first quantization by adding a new label to the states. Thus 
\begin{equation}
    \ket{\psi_{\nu,m}} =\frac{1}{\sqrt{N!}} \sum_i \eta_i \ket{\psi_{\nu,m,i}}
\label{eq:fqbasis}
\end{equation}
where the new index $i$ runs over the $N!{}$ different orientations of the particle labels and $\eta_i =\pm 1$ is the corresponding phase factor,
\begin{align*}
    \ket{\psi_{2,1}} &\equiv |110100\rangle \\
                     &=\frac{1}{\sqrt{6}}\left(|1_{1}1_{2}01_{3}00\rangle+|1_{3}1_{1}01_{2}00\rangle+|1_{2}1_{3}01_{1}00\rangle \right. \\
                     & \left.\quad - |1_{2}1_{1}01_{3}00\rangle-|1_{3}1_{2}01_{1}00\rangle-|1_{1}1_{3}01_{2}00\rangle\right) 
\end{align*}
where the subscripts are particles labels and we use the usual sign convention based on their permutations.   

Now we can partition the particles into two sets, containing $n=1$ particle, say the particle with the label $1$, and the remaining $N-n=2$ particles with  labels $2$ and $3$. Any $N=3$ particle state can be written as a tensor product from the two subsets. For example, $\ket{1_{1}1_{2}01_{3}00}=\ket{1_{1}00000}\otimes \ket{01_{2}01_{3}00}$. Performing this decomposition entails finding the coefficients $b_{\nu,\nu^\prime,m,m^\prime,i,i^\prime}$, such that the state can be expanded as
\begin{equation*}
\ket{\Psi_{0}} =\sum_{\nu,\nu^\prime}\sum_{m,m^\prime}\sum_{i,i^\prime}
b_{\nu,\nu^\prime,m,m^\prime,i,i^\prime} \ket{\psi^{(n)}_{\nu,m,i}}\ket{\psi^{(N-n)}_{\nu^\prime,m^\prime,i^\prime}}, 
\end{equation*}
where $\ket{\psi^{(n)}_{\nu,m,i}}$ and $\ket{\psi^{(N-n)}_{\nu^\prime,m^\prime,i^\prime}}$ represent the first quantization basis states for the two groups of particles, respectively. 
The resulting Schmidt decomposition matrix of the state $\ket{\Psi_0}$ is given by
\begin{equation}
\mathsf{G}^{(n)} = \sum_{\nu,\nu^\prime}\sum_{m,m^\prime}\sum_{i,i^\prime}
b_{\nu,\nu^\prime,m,m^\prime,i,i^\prime} \ket{\psi^{(n)}_{\nu,m,i}}\bra{\psi^{(N-n)}_{\nu^\prime,m^\prime,i^\prime}}.
\label{eq:Gn}
\end{equation}

In a previous work \cite{Barghathi:2017fv}, we have shown that the spectrum of a $n$-body reduced density matrix $\rho_n$ can be obtained by considering a smaller matrix $\tilde{\mathsf{G}}^{(n)}$ that contains only $\frac{1}{n!(N-n)!}$ of the number of elements in $\mathsf{G}^{(n)}$. The matrix $\tilde{\mathsf{G}}^{(n)}$ is obtained by choosing a specific orientation of the particles labels in any of the subsets (i.e.\ increasing order) and keeping track of the overall phase (signs) of a $N$ particle configuration by considering the relative orientation of the particles from the two sets.

\subsection{Application of translational symmetry to the occupation subsets} 

We now consider the effect of translational symmetry on the particle sub-group occupation states.  The number of possible cycles depends on the number of particles in the group and we can suppress the explicit particle labels (e.g. $1_1,1_2,1_3$) by fixing the orientation such that they are always in increasing order when written from left to right in a subgroup.  We then use a primed notation to distinguish states in the group with $n=1$ where there is only one translational cycle with six elements:
%
\begin{align}
    \ket*{\psi^{(n)}_{1,1}} &= T^{0}\ket*{\psi^{(n)}_{1,1}}=\ket{1^{\prime}00000}\nonumber\\
   \ket*{\psi^{(n)}_{1,2}} &= T^{1}\ket*{\psi^{(n)}_{1,1}}= \ket{01^{\prime}0000}\nonumber\\
   \ket*{\psi^{(n)}_{1,3}} &= T^{2}\ket*{\psi^{(n)}_{1,1}}= \ket{001^{\prime}000}\nonumber\\
   \ket*{\psi^{(n)}_{1,4}} &= T^{3}\ket*{\psi^{(n)}_{1,1}}= \ket{0001^{\prime}00}\nonumber\\
   \ket*{\psi^{(n)}_{1,5}} &= T^{4}\ket*{\psi^{(n)}_{1,1}}= \ket{00001^{\prime}0}\nonumber\\
   \ket*{\psi^{(n)}_{1,6}} &= T^{5}\ket*{\psi^{(n)}_{1,1}}= \ket{000001^{\prime}}
\label{eq:1Cycle1}
\end{align}
from the $\tbinom{L}{N-n}=\tbinom{6}{2} = 15$ occupation states in the $N-n=2$ group. The latter can be decomposed into three translational cycles as follows:
\begin{align*}
   \ket*{\psi^{(N-n)}_{1,1}} &= T^{0}\ket*{\psi^{(N-n)}_{1,1}}=\ket{110000} \\ 
   \ket*{\psi^{(N-n)}_{1,2}} &= T^{1}\ket*{\psi^{(N-n)}_{1,1}}=\ket{011000} \\
   \ket*{\psi^{(N-n)}_{1,3}} &= T^{2}\ket*{\psi^{(N-n)}_{1,1}}=\ket{001100} \\ 
   \ket*{\psi^{(N-n)}_{1,4}} &= T^{3}\ket*{\psi^{(N-n)}_{1,1}}=\ket{000110} \\ 
   \ket*{\psi^{(N-n)}_{1,5}} &= T^{4}\ket*{\psi^{(N-n)}_{1,1}}=\ket{000011} \\
   \ket*{\psi^{(N-n)}_{1,6}} &= T^{5}\ket*{\psi^{(N-n)}_{1,1}}=\ket{100001} \\
\end{align*}

\begin{align*}
   \ket*{\psi^{(N-n)}_{2,1}} &= T^{0}\ket*{\psi^{(N-n)}_{2,1}}=\ket{101000} \\
   \ket*{\psi^{(N-n)}_{2,2}} &= T^{1}\ket*{\psi^{(N-n)}_{2,1}}=\ket{010100} \\
   \ket*{\psi^{(N-n)}_{2,3}} &= T^{2}\ket*{\psi^{(N-n)}_{2,1}}=\ket{001010} \\
   \ket*{\psi^{(N-n)}_{2,4}} &= T^{3}\ket*{\psi^{(N-n)}_{2,1}}=\ket{000101} \\
   \ket*{\psi^{(N-n)}_{2,5}} &= T^{4}\ket*{\psi^{(N-n)}_{2,1}}=\ket{100010} \\
   \ket*{\psi^{(N-n)}_{2,6}} &= T^{5}\ket*{\psi^{(N-n)}_{2,1}}=\ket{010001}
\end{align*}
\begin{align}
    \ket*{\psi^{(N-n)}_{3,1}} &= T^{0}\ket*{\psi^{(N-n)}_{3,1}}=T^{3}\ket*{\psi^{(N-n)}_{3,1}}=\ket{100100}\nonumber\\ 
    \ket*{\psi^{(N-n)}_{3,2}} &= T^{1}\ket*{\psi^{(N-n)}_{3,1}}=T^{4}\ket*{\psi^{(N-n)}_{3,1}}=\ket{010010}\nonumber\\
    \ket*{\psi^{(N-n)}_{3,3}} &= T^{2}\ket*{\psi^{(N-n)}_{3,1}}=T^{5}\ket*{\psi^{(N-n)}_{3,1}}=\ket{001001}
\label{eq:2Cycle3}.
\end{align}

We have now exposed enough structure to express the Schmidt decomposition matrix $\tilde{\mathsf{G}}^{(n)}$ as being composed of three sub-matrices 
\begin{equation}
    \tilde{\mathsf{G}}^{(n)}=
\begin{bmatrix}
    \mathsf{A}_{1,1} & \mathsf{A}_{1,2}&\mathsf{A}_{1,3}
\end{bmatrix}
\label{eq:GA},
\end{equation}
where $\mathsf{A}_{\nu^\prime,\nu}=\sum_{m^\prime,m}c_{\nu^\prime,\nu,m^\prime,m} \ket*{\psi^{(n)}_{\nu^\prime,m^\prime}} \bra*{\psi^{(N-n)}_{\nu,m}}$ 
and the coefficients $c_{\nu^\prime,\nu,m^\prime,m}$ can be read off from Eq.~(\ref{eq:GST}) combined with Eq.~(\ref{eq:fqbasis}) to yield
\begin{equation}
    \mathsf{A}_{1,1}=
\begin{bmatrix}
0&\bar{a}_1&\bar{a}_3&\bar{a}_2&\bar{a}_1&0\\
0&0&\bar{a}_1&\bar{a}_3&\bar{a}_2&-\bar{a}_1\\
\bar{a}_1&0&0&\bar{a}_1&\bar{a}_3&-\bar{a}_2\\
\bar{a}_2&\bar{a}_1&0&0&\bar{a}_1&-\bar{a}_3\\
\bar{a}_3 &\bar{a}_2&\bar{a}_1&0&0&-\bar{a}_1\\
\bar{a}_1&\bar{a}_3&\bar{a}_2&\bar{a}_1&0&0
\end{bmatrix}
\label{eq:A11},
\end{equation}
\begin{equation}
    \mathsf{A}_{1,2}=
\begin{bmatrix}
0&\bar{a}_2&\bar{a}_4&\bar{a}_3&0&\bar{a}_1\\
-\bar{a}_1&0&\bar{a}_2&\bar{a}_4&-\bar{a}_3&0\\
0&-\bar{a}_1&0&\bar{a}_2&-\bar{a}_4&-\bar{a}_3\\
\bar{a}_3&0&-\bar{a}_1&0&-\bar{a}_2&-\bar{a}_4\\
\bar{a}_4 &\bar{a}_3&0&-\bar{a}_1&0&-\bar{a}_2\\
\bar{a}_2&\bar{a}_4&\bar{a}_3&0&\bar{a}_1&0
\end{bmatrix}
\label{eq:A12},
\end{equation}
\begin{equation}
    \mathsf{A}_{1,3}=
\begin{bmatrix}
0&\bar{a}_3&\bar{a}_2\\
-\bar{a}_2&0&\bar{a}_3\\
-\bar{a}_3&-\bar{a}_2&0\\
0&-\bar{a}_3&-\bar{a}_2\\
\bar{a}_2&0&-\bar{a}_3\\
\bar{a}_3&\bar{a}_2&0
\end{bmatrix}
\label{eq:A13},
\end{equation}
with $\bar{a}_1=a_1/6$, $\bar{a}_2=a_2/6$, $\bar{a}_3=a_3/6$ and $\bar{a}_4=a_4/\sqrt{12}$. To understand how these are actually obtained it is useful to consider a specific example for the element $[\mathsf{A}_{1,2}]_{1,3}$ which is the coefficient corresponding to $\ket*{\psi^{(n)}_{1,1}}\bra*{\psi^{(N-n)}_{2,3}}=\ket{1^{\prime}00000}\bra{001010}$, which comes from the decomposition of the $N$-particle state $\ket{1^{\prime}01010}$. Re-introducing the particle labels $\ket{1_101_201_30}$, we note the orientation has a positive phase and it appears in the ground state only through $\ket{\psi_{4,1}}$ with a factor of $1/\sqrt{6}$ and the latter has a unique contribution through $\ket{\phi_{4,0}}$ and thus the targeted coefficient is $\bar{a}_4=a_4/(\sqrt{2}\sqrt{6})$.  Similarly, starting from this position and moving one step in the diagonal direction results in shifting all the particles one site to the right, and thus we get $[\mathsf{A}_{1,2}]_{2,4}$ as the coefficient of $\ket{01_101_201_3}$ which is also $\bar{a}_4$ due to the translational symmetry of the ground state. However, if we proceed along the same diagonal and evaluate the coefficient $[\mathsf{A}_{1,2}]_{3,5}$ we get $-\bar{a}_4$ as it corresponds to $\ket{1_201_101_30}$ which has a negative phase as particle $2$ in group $2$ has wrapped around the boundary. The appearance of this minus sign is somewhat spurious, and arises from the chosen first-quantized labelling scheme of particles in the subgroups in increasing order.  This can be understood by considering Eq.~(\ref{eq:2Cycle3}) where the translational symmetry such that $T^6 = 1$ is arising from true indistinguishability of the particles.  Such signs are always attached to either a full row or column and we note that if we were to multiply columns $5$ and $6$ of the matrix $\mathsf{A}_{1,2}$ by $-1$ then the resulting matrix is periodic. Similarly, multiplying the sixth column of $\mathsf{A}_{1,1}$ by $-1$ results in a periodic matrix. Also, the matrix $A_{1,3}$ is periodic in the vertical direction (rows), while its antiperiodic in the horizontal direction (columns).

In general, if we account for the negative signs that are attached to columns and/or rows, the resulting $\mathsf{A}$-matrices are either periodic, antiperiodic or mixed, depending on the relationship between the number of particles in each subgroup and the number of elements in the symmetry cycles involved.   The spatial symmetries can be determined by computing the parity of the product $n(N-n)M_{\nu^\prime}^{(n)}/L$ for rows and $n(N-n)M_{\nu}^{(N-n)}/L$ for columns with even/odd corresponding to periodic/anti-periodic and $M_{\nu^\prime}^{(n)}$ is the number of elements in the translational cycle $\nu^\prime$ corresponding to the $n$-particle group.

Based on this analysis, we can build unitary transformations to simplify the matrix $\tilde{\mathsf{G}}_{(n)}$. We begin by defining unitary operators that diagonalize the matrices $\mathsf{A}_{\nu^\prime,\nu}$.  Starting with $\mathsf{A}_{1,1}$, we first account for the row/column spurious signs which can be dealt with via
the unitary operator $\mathsf{P}_1$, with $[\mathsf{P}_1]_{m^\prime,m}=0$ for $m^\prime\ne m$,  $[\mathsf{P}_1]_{6,6}=-1$ and $[\mathsf{P}_1]_{m,m}=1$ for $m\le6$. The matrix $\mathsf{A}_{1,1}\mathsf{P}^{\dagger}_1$, is then fully periodic and can be diagonalized with the periodic square Fourier transform matrix $\mathsf{F_\nu}$ as 
\begin{equation}
    \mathsf{D}_{1,1}=\mathsf{F}_1\mathsf{A}_{1,1}\mathsf{P}^\dagger_1\mathsf{F}^\dagger_1
\label{eq:D1},
\end{equation}
where 
\begin{equation}
    [\mathsf{F}_\nu]_{m^\prime,m}=\frac{1}{\sqrt{M_\nu}}\mathrm{e}^{-\imath 2\pi(m^\prime-1)(m-1)/M_\nu}. 
\label{eq:Fp}
\end{equation}
In the same fashion, we can diagonalize $\mathsf{A}_{1,2}$ as
\begin{equation}
    \mathsf{D}_{1,2}=\mathsf{F}_1\mathsf{A}_{1,2}\mathsf{P}^\dagger_2\mathsf{F}^\dagger_2
\label{eq:D2},
\end{equation}
where we have accounted for the extra signs via $\mathsf{P}_2$ which has $[\mathsf{P}_2]_{m^\prime,m}=0$ for $m^\prime\ne m$,  $[\mathsf{P}_2]_{5,5}=[\mathsf{P}_1]_{6,6}=-1$ and $[\mathsf{P}_2]_{m,m}=1$ for $m\le4$.  
Finally, the rectangular matrix with mixed periodicity/anti-periodicity $\mathsf{A}_{1,3}$ can be diagonalized as
\begin{equation}
    \mathsf{D}_{1,3}=\mathsf{F}_1\mathsf{A}_{1,3}\mathsf{\tilde{F}}^\dagger_3
\label{eq:D3},
\end{equation}
where the anti-periodic Fourier matrix is
\begin{equation}
    [\mathsf{\tilde{F}}_\nu]_{m^\prime,m}=\frac{1}{\sqrt{M_\nu}}\mathrm{e}^{-\imath 2\pi(m-1)(m^\prime-1/2)/M_\nu} 
\label{eq:Fa}
\end{equation}
with $M_3=3$ corresponding to the number of states in the 3rd cycle for the $N-n$ group of particles.  The matrix 
\begin{equation}
    \mathsf{D}=
\begin{bmatrix}
    \mathsf{D}_{1,1}& \mathsf{D}_{1,2}& \mathsf{D}_{1,3}
\end{bmatrix}
\label{eq:D},
\end{equation}
can be obtained directly from $\tilde{\mathsf{G}}^{(n)}$ via $\mathsf{D}=\mathsf{U}\tilde{\mathsf{G}}^{(n)}\mathsf{V}^\dagger$, where
\begin{equation}
    \mathsf{U}=\mathsf{F}_1
\label{eq:P_l},
\end{equation}
and
\begin{equation}
    \mathsf{V}^\dagger=
\begin{bmatrix}
    \mathsf{P}_1^\dagger\mathsf{F}^\dagger_1&0&0\\
    0&\mathsf{P}^\dagger_2\mathsf{F}^\dagger_2&0\\
    0&0&\mathsf{\tilde{F}}^\dagger_3
\end{bmatrix}
\label{eq:P_r}\,.
\end{equation}
\setcounter{MaxMatrixCols}{20}
We now arrive at the explicit form of the $\mathsf{D}$ matrix
\begin{widetext}
\begin{equation}
    \mathsf{D}=
\begingroup
\setlength\arraycolsep{1.5pt}
\begin{bmatrix}
    [\mathsf{D}_{1,1}]_1&0&0&0&0&0&[\mathsf{D}_{1,2}]_1&0&0&0&0&0&0&0&0\\
    0&[\mathsf{D}_{1,1}]_2&0&0&0&0&0&[\mathsf{D}_{1,2}]_2&0&0&0&0&[\mathsf{D}_{1,3}]_1&0&0\\
    0&0&[\mathsf{D}_{1,1}]_3&0&0&0&0&0&[\mathsf{D}_{1,2}]_3&0&0&0&0&0&0\\
    0&0&0&[\mathsf{D}_{1,1}]_4&0&0&0&0&0&[\mathsf{D}_{1,2}]_4&0&0&0&[\mathsf{D}_{1,3}]_2&0\\
    0&0&0&0&[\mathsf{D}_{1,1}]_5&0&0&0&0&0&[\mathsf{D}_{1,2}]_5&0&0&0&0\\
    0&0&0&0&0&[\mathsf{D}_{1,1}]_6&0&0&0&0&0&[\mathsf{D}_{1,2}]_6&0&0&[\mathsf{D}_{1,3}]_3
\end{bmatrix}
\endgroup
\label{eq:Dexplicitly}
\end{equation}
which can be put in a block diagonal form by a rearrangement of the columns and rows (in this example, we only rearrange the columns) via a final unitary transformation that exchanges the basis of $\mathsf{D}$. This leads to
\begin{equation}
    \tilde{\mathsf{D}}=
\begingroup
\setlength\arraycolsep{1.5pt}
\begin{bmatrix}
    [\mathsf{D}_{1,1}]_1&[\mathsf{D}_{1,2}]_1&0&0&0&0&0&0&0&0&0&0&0&0&0\\
    0&0&[\mathsf{D}_{1,1}]_2&[\mathsf{D}_{1,2}]_2&[\mathsf{D}_{1,3}]_1&0&0&0&0&0&0&0&0&0&0\\
    0&0&0&0&0&[\mathsf{D}_{1,1}]_3&[\mathsf{D}_{1,2}]_3&0&0&0&0&0&0&0&0\\
    0&0&0&0&0&0&0&[\mathsf{D}_{1,1}]_4&[\mathsf{D}_{1,2}]_4&[\mathsf{D}_{1,3}]_2&0&0&0&0&0\\
    0&0&0&0&0&0&0&0&0&0&[\mathsf{D}_{1,1}]_5&[\mathsf{D}_{1,2}]_5&0&0&0\\
    0&0&0&0&0&0&0&0&0&0&0&0&[\mathsf{D}_{1,1}]_6&[\mathsf{D}_{1,2}]_6&[\mathsf{D}_{1,3}]_3
\end{bmatrix}
\endgroup
\label{eq:tiledD}.
\end{equation}
\end{widetext}
A singular value decomposition of $\mathsf{D}$ can be performed by obtaining the singular values of each of the six blocks. In this example we find:
$d_1=1/\sqrt{6}$, $d_2=1/\sqrt{6}$, $d_3=0$, $d_4=0$, $d_5=0$ and $d_6=1/\sqrt{6}$.
The resulting eigenvalues of the $n$-body reduced density matrix $\rho_n$ are \cite{Barghathi:2017fv}: 
\begin{equation}
\lambda_k=n!(N-n)!d_k^2\, ,
\end{equation}
thus:
\begin{align}
    \lambda_1 &= \frac{1}{3} & \lambda_2 &=\frac{1}{3} & \lambda_3 & =0 & \lambda_4& =0 & \lambda_5 & =0  & \lambda_6 & = \frac{1}{3}.
\label{eq:lambda}
\end{align}

This efficient approach can be compared with the brute-force construction of the $n$-particle reduced density matrix for this case using no particle subgroup symmetries which yields
\begin{equation}
\rho_n =
\begin{bmatrix}
 \alpha_1 & \alpha_2 & 0 & \alpha_3 & 0 & \alpha_2 \\
 \alpha_2 & \alpha_1 & \alpha_2 & 0 & \alpha_3 & 0 \\
 0 & \alpha_2 & \alpha_1 & \alpha_2 & 0 & \alpha_3 \\
 \alpha_3 & 0 & \alpha_2 & \alpha_1 & \alpha_2 & 0 \\
 0 & \alpha_3 & 0 & \alpha_2 & \alpha_1 & \alpha_2 \\
 \alpha_2 & 0 & \alpha_3 & 0 & \alpha_2 & \alpha_1 \\
\end{bmatrix}
\label{eq:rhonmat}
\end{equation}
where
\begin{align}
    \alpha_1 &= \frac{1}{6} \qty(a_1^2 + a_2^2 + a_3^2 + a_4^2) \nonumber \\
    \alpha_2 &= \frac{1}{18} \qty[a_1 (a_2+a_3)+2 a_2 a_3+\sqrt{3} a_2 a_4+\sqrt{3} a_3 a_4] \nonumber \\
    \alpha_3 &= \frac{1}{18} \left(2 a_1^2-2 \sqrt{3} a_1 a_4-a_2^2-a_3^2\right).
\label{eq:coeffs}
\end{align}
The eigenvalues can be easily confirmed to yield $\lambda_k$ but here we must diagonalize one  $\binom{L}{n} \times \binom{L}{n}$ matrix as opposed to $L$ considerably smaller matrices whose maximal linear dimension can be reduced by a factor up to $\max[n!,(N-n)!]L$.  \\

The full implementation of these particle subgroup translational symmetries (with details in the released code \cite{repo}) has allowed us to compute the post-quench dynamics of particle entanglement entropies for reduced density matrices with $n=\lfloor N/2 \rfloor = 6$ for systems up to $L=26$ sites and $N=13$ fermions at half filling for long times $tJ = 100$.

\bibliography{refs}

\begin{thebibliography}{51}%
\makeatletter
\providecommand \@ifxundefined [1]{%
 \@ifx{#1\undefined}
}%
\providecommand \@ifnum [1]{%
 \ifnum #1\expandafter \@firstoftwo
 \else \expandafter \@secondoftwo
 \fi
}%
\providecommand \@ifx [1]{%
 \ifx #1\expandafter \@firstoftwo
 \else \expandafter \@secondoftwo
 \fi
}%
\providecommand \natexlab [1]{#1}%
\providecommand \enquote  [1]{``#1''}%
\providecommand \bibnamefont  [1]{#1}%
\providecommand \bibfnamefont [1]{#1}%
\providecommand \citenamefont [1]{#1}%
\providecommand \href@noop [0]{\@secondoftwo}%
\providecommand \href [0]{\begingroup \@sanitize@url \@href}%
\providecommand \@href[1]{\@@startlink{#1}\@@href}%
\providecommand \@@href[1]{\endgroup#1\@@endlink}%
\providecommand \@sanitize@url [0]{\catcode `\\12\catcode `\$12\catcode
  `\&12\catcode `\#12\catcode `\^12\catcode `\_12\catcode `\%12\relax}%
\providecommand \@@startlink[1]{}%
\providecommand \@@endlink[0]{}%
\providecommand \url  [0]{\begingroup\@sanitize@url \@url }%
\providecommand \@url [1]{\endgroup\@href {#1}{\urlprefix }}%
\providecommand \urlprefix  [0]{URL }%
\providecommand \Eprint [0]{\href }%
\providecommand \doibase [0]{http://dx.doi.org/}%
\providecommand \selectlanguage [0]{\@gobble}%
\providecommand \bibinfo  [0]{\@secondoftwo}%
\providecommand \bibfield  [0]{\@secondoftwo}%
\providecommand \translation [1]{[#1]}%
\providecommand \BibitemOpen [0]{}%
\providecommand \bibitemStop [0]{}%
\providecommand \bibitemNoStop [0]{.\EOS\space}%
\providecommand \EOS [0]{\spacefactor3000\relax}%
\providecommand \BibitemShut  [1]{\csname bibitem#1\endcsname}%
\let\auto@bib@innerbib\@empty
\bibitem [{\citenamefont {Srednicki}(1994)}]{Srednicki:1994xg}%
  \BibitemOpen
  \bibfield  {author} {\bibinfo {author} {\bibfnamefont {Mark}\ \bibnamefont
  {Srednicki}},\ }\bibfield  {title} {\enquote {\bibinfo {title} {{C}haos and
  quantum thermalization},}\ }\href {\doibase 10.1103/physreve.50.888}
  {\bibfield  {journal} {\bibinfo  {journal} {Phys. Rev. E}\ }\textbf {\bibinfo
  {volume} {50}},\ \bibinfo {pages} {888} (\bibinfo {year} {1994})}\BibitemShut
  {NoStop}%
\bibitem [{\citenamefont {Rigol}\ \emph {et~al.}(2008)\citenamefont {Rigol},
  \citenamefont {Dunjko},\ and\ \citenamefont {Olshanii}}]{Rigol:2008sl}%
  \BibitemOpen
  \bibfield  {author} {\bibinfo {author} {\bibfnamefont {Marcos}\ \bibnamefont
  {Rigol}}, \bibinfo {author} {\bibfnamefont {Vanja}\ \bibnamefont {Dunjko}}, \
  and\ \bibinfo {author} {\bibfnamefont {Maxim}\ \bibnamefont {Olshanii}},\
  }\bibfield  {title} {\enquote {\bibinfo {title} {{T}hermalization and its
  mechanism for generic isolated quantum systems},}\ }\href {\doibase
  10.1038/nature06838} {\bibfield  {journal} {\bibinfo  {journal} {Nature}\
  }\textbf {\bibinfo {volume} {452}},\ \bibinfo {pages} {854} (\bibinfo {year}
  {2008})}\BibitemShut {NoStop}%
\bibitem [{\citenamefont {Calabrese}\ and\ \citenamefont
  {Cardy}(2005)}]{Calabrese:2005oi}%
  \BibitemOpen
  \bibfield  {author} {\bibinfo {author} {\bibfnamefont {Pasquale}\
  \bibnamefont {Calabrese}}\ and\ \bibinfo {author} {\bibfnamefont {John}\
  \bibnamefont {Cardy}},\ }\bibfield  {title} {\enquote {\bibinfo {title}
  {{E}volution of entanglement entropy in one-dimensional systems},}\ }\href
  {\doibase 10.1088/1742-5468/2005/04/p04010} {\bibfield  {journal} {\bibinfo
  {journal} {J. Stat. Mech.: Theor. Exp.}\ }\textbf {\bibinfo {volume}
  {2005}},\ \bibinfo {pages} {P04010} (\bibinfo {year} {2005})}\BibitemShut
  {NoStop}%
\bibitem [{\citenamefont {Polkovnikov}\ \emph {et~al.}(2011)\citenamefont
  {Polkovnikov}, \citenamefont {Sengupta}, \citenamefont {Silva},\ and\
  \citenamefont {Vengalattore}}]{Polkovnikov:2011gg}%
  \BibitemOpen
  \bibfield  {author} {\bibinfo {author} {\bibfnamefont {Anatoli}\ \bibnamefont
  {Polkovnikov}}, \bibinfo {author} {\bibfnamefont {Krishnendu}\ \bibnamefont
  {Sengupta}}, \bibinfo {author} {\bibfnamefont {Alessandro}\ \bibnamefont
  {Silva}}, \ and\ \bibinfo {author} {\bibfnamefont {Mukund}\ \bibnamefont
  {Vengalattore}},\ }\bibfield  {title} {\enquote {\bibinfo {title}
  {{C}olloquium: {N}onequilibrium dynamics of closed interacting quantum
  systems},}\ }\href {\doibase 10.1103/revmodphys.83.863} {\bibfield  {journal}
  {\bibinfo  {journal} {Rev. Mod. Phys.}\ }\textbf {\bibinfo {volume} {83}},\
  \bibinfo {pages} {863} (\bibinfo {year} {2011})}\BibitemShut {NoStop}%
\bibitem [{\citenamefont {D'Alessio}\ \emph {et~al.}(2016)\citenamefont
  {D'Alessio}, \citenamefont {Kafri}, \citenamefont {Polkovnikov},\ and\
  \citenamefont {Rigol}}]{DAlessio:2016rr}%
  \BibitemOpen
  \bibfield  {author} {\bibinfo {author} {\bibfnamefont {Luca}\ \bibnamefont
  {D'Alessio}}, \bibinfo {author} {\bibfnamefont {Yariv}\ \bibnamefont
  {Kafri}}, \bibinfo {author} {\bibfnamefont {Anatoli}\ \bibnamefont
  {Polkovnikov}}, \ and\ \bibinfo {author} {\bibfnamefont {Marcos}\
  \bibnamefont {Rigol}},\ }\bibfield  {title} {\enquote {\bibinfo {title} {From
  quantum chaos and eigenstate thermalization to statistical mechanics and
  thermodynamics},}\ }\href {\doibase 10.1080/00018732.2016.1198134} {\bibfield
   {journal} {\bibinfo  {journal} {Adv. in Phys.}\ }\textbf {\bibinfo {volume}
  {65}},\ \bibinfo {pages} {239} (\bibinfo {year} {2016})}\BibitemShut
  {NoStop}%
\bibitem [{\citenamefont {Alba}\ and\ \citenamefont
  {Calabrese}(2017)}]{Alba:2017ph}%
  \BibitemOpen
  \bibfield  {author} {\bibinfo {author} {\bibfnamefont {Vincenzo}\
  \bibnamefont {Alba}}\ and\ \bibinfo {author} {\bibfnamefont {Pasquale}\
  \bibnamefont {Calabrese}},\ }\bibfield  {title} {\enquote {\bibinfo {title}
  {{E}ntanglement and thermodynamics after a quantum quench in integrable
  systems},}\ }\href {\doibase 10.1073/pnas.1703516114} {\bibfield  {journal}
  {\bibinfo  {journal} {Proc. Nat. Acad. Sci.}\ }\textbf {\bibinfo {volume}
  {114}},\ \bibinfo {pages} {7947} (\bibinfo {year} {2017})}\BibitemShut
  {NoStop}%
\bibitem [{\citenamefont {Kaufman}\ \emph {et~al.}(2016)\citenamefont
  {Kaufman}, \citenamefont {Tai}, \citenamefont {Lukin}, \citenamefont
  {Rispoli}, \citenamefont {Schittko}, \citenamefont {Preiss},\ and\
  \citenamefont {Greiner}}]{Kaufman:2016ep}%
  \BibitemOpen
  \bibfield  {author} {\bibinfo {author} {\bibfnamefont {Adam~M.}\ \bibnamefont
  {Kaufman}}, \bibinfo {author} {\bibfnamefont {M.~Eric}\ \bibnamefont {Tai}},
  \bibinfo {author} {\bibfnamefont {Alexander}\ \bibnamefont {Lukin}}, \bibinfo
  {author} {\bibfnamefont {Matthew}\ \bibnamefont {Rispoli}}, \bibinfo {author}
  {\bibfnamefont {Robert}\ \bibnamefont {Schittko}}, \bibinfo {author}
  {\bibfnamefont {Philipp~M.}\ \bibnamefont {Preiss}}, \ and\ \bibinfo {author}
  {\bibfnamefont {Markus}\ \bibnamefont {Greiner}},\ }\bibfield  {title}
  {\enquote {\bibinfo {title} {{Q}uantum thermalization through entanglement in
  an isolated many-body system},}\ }\href {\doibase 10.1126/science.aaf6725}
  {\bibfield  {journal} {\bibinfo  {journal} {Science}\ }\textbf {\bibinfo
  {volume} {353}},\ \bibinfo {pages} {794} (\bibinfo {year}
  {2016})}\BibitemShut {NoStop}%
\bibitem [{\citenamefont {Lukin}\ \emph {et~al.}(2019)\citenamefont {Lukin},
  \citenamefont {Rispoli}, \citenamefont {Schittko}, \citenamefont {Tai},
  \citenamefont {Kaufman}, \citenamefont {Choi}, \citenamefont {Khemani},
  \citenamefont {L{\'e}onard},\ and\ \citenamefont {Greiner}}]{Lukin:2018wg}%
  \BibitemOpen
  \bibfield  {author} {\bibinfo {author} {\bibfnamefont {Alexander}\
  \bibnamefont {Lukin}}, \bibinfo {author} {\bibfnamefont {Matthew}\
  \bibnamefont {Rispoli}}, \bibinfo {author} {\bibfnamefont {Robert}\
  \bibnamefont {Schittko}}, \bibinfo {author} {\bibfnamefont {M.~Eric}\
  \bibnamefont {Tai}}, \bibinfo {author} {\bibfnamefont {Adam~M.}\ \bibnamefont
  {Kaufman}}, \bibinfo {author} {\bibfnamefont {Soonwon}\ \bibnamefont {Choi}},
  \bibinfo {author} {\bibfnamefont {Vedika}\ \bibnamefont {Khemani}}, \bibinfo
  {author} {\bibfnamefont {Julian}\ \bibnamefont {L{\'e}onard}}, \ and\
  \bibinfo {author} {\bibfnamefont {Markus}\ \bibnamefont {Greiner}},\
  }\bibfield  {title} {\enquote {\bibinfo {title} {Probing entanglement in a
  many-body-localized system},}\ }\href {\doibase 10.1126/science.aau0818}
  {\bibfield  {journal} {\bibinfo  {journal} {Science}\ }\textbf {\bibinfo
  {volume} {364}},\ \bibinfo {pages} {256} (\bibinfo {year}
  {2019})}\BibitemShut {NoStop}%
\bibitem [{\citenamefont {Kinoshita}\ \emph {et~al.}(2006)\citenamefont
  {Kinoshita}, \citenamefont {Wenger},\ and\ \citenamefont
  {Weiss}}]{Kinoshita:2006yz}%
  \BibitemOpen
  \bibfield  {author} {\bibinfo {author} {\bibfnamefont {Toshiya}\ \bibnamefont
  {Kinoshita}}, \bibinfo {author} {\bibfnamefont {Trevor}\ \bibnamefont
  {Wenger}}, \ and\ \bibinfo {author} {\bibfnamefont {David~S.}\ \bibnamefont
  {Weiss}},\ }\bibfield  {title} {\enquote {\bibinfo {title} {{A} quantum
  {N}ewton's cradle},}\ }\href {\doibase 10.1038/nature04693} {\bibfield
  {journal} {\bibinfo  {journal} {Nature}\ }\textbf {\bibinfo {volume} {440}},\
  \bibinfo {pages} {900} (\bibinfo {year} {2006})}\BibitemShut {NoStop}%
\bibitem [{\citenamefont {Langen}\ \emph {et~al.}(2013)\citenamefont {Langen},
  \citenamefont {Geiger}, \citenamefont {Kuhnert}, \citenamefont {Rauer},\ and\
  \citenamefont {Schmiedmayer}}]{Langen:2013ls}%
  \BibitemOpen
  \bibfield  {author} {\bibinfo {author} {\bibfnamefont {T.}~\bibnamefont
  {Langen}}, \bibinfo {author} {\bibfnamefont {R.}~\bibnamefont {Geiger}},
  \bibinfo {author} {\bibfnamefont {M.}~\bibnamefont {Kuhnert}}, \bibinfo
  {author} {\bibfnamefont {B.}~\bibnamefont {Rauer}}, \ and\ \bibinfo {author}
  {\bibfnamefont {J.}~\bibnamefont {Schmiedmayer}},\ }\bibfield  {title}
  {\enquote {\bibinfo {title} {{L}ocal emergence of thermal correlations in an
  isolated quantum many-body system},}\ }\href {\doibase 10.1038/nphys2739}
  {\bibfield  {journal} {\bibinfo  {journal} {Nature Physics}\ }\textbf
  {\bibinfo {volume} {9}},\ \bibinfo {pages} {640} (\bibinfo {year}
  {2013})}\BibitemShut {NoStop}%
\bibitem [{\citenamefont {Haque}\ \emph {et~al.}(2007)\citenamefont {Haque},
  \citenamefont {Zozulya},\ and\ \citenamefont {Schoutens}}]{Haque:2007il}%
  \BibitemOpen
  \bibfield  {author} {\bibinfo {author} {\bibfnamefont {Masudul}\ \bibnamefont
  {Haque}}, \bibinfo {author} {\bibfnamefont {Oleksandr}\ \bibnamefont
  {Zozulya}}, \ and\ \bibinfo {author} {\bibfnamefont {Kareljan}\ \bibnamefont
  {Schoutens}},\ }\bibfield  {title} {\enquote {\bibinfo {title} {{Entanglement
  Entropy in Fermionic Laughlin States}},}\ }\href
  {https://doi.org/10.1103/PhysRevLett.98.060401} {\bibfield  {journal}
  {\bibinfo  {journal} {Phys. Rev. Lett.}\ }\textbf {\bibinfo {volume} {98}},\
  \bibinfo {pages} {060401} (\bibinfo {year} {2007})}\BibitemShut {NoStop}%
\bibitem [{\citenamefont {Zozulya}\ \emph {et~al.}(2007)\citenamefont
  {Zozulya}, \citenamefont {Haque}, \citenamefont {Schoutens},\ and\
  \citenamefont {Rezayi}}]{Zozulya:2007jw}%
  \BibitemOpen
  \bibfield  {author} {\bibinfo {author} {\bibfnamefont {O.S.}\ \bibnamefont
  {Zozulya}}, \bibinfo {author} {\bibfnamefont {M}~\bibnamefont {Haque}},
  \bibinfo {author} {\bibfnamefont {K}~\bibnamefont {Schoutens}}, \ and\
  \bibinfo {author} {\bibfnamefont {E.H.}\ \bibnamefont {Rezayi}},\ }\bibfield
  {title} {\enquote {\bibinfo {title} {{Bipartite entanglement entropy in
  fractional quantum Hall states}},}\ }\href
  {https://doi.org/10.1103/PhysRevB.76.125310} {\bibfield  {journal} {\bibinfo
  {journal} {Phys. Rev. B}\ }\textbf {\bibinfo {volume} {76}},\ \bibinfo
  {pages} {125310} (\bibinfo {year} {2007})}\BibitemShut {NoStop}%
\bibitem [{\citenamefont {Zozulya}\ \emph {et~al.}(2008)\citenamefont
  {Zozulya}, \citenamefont {Haque},\ and\ \citenamefont
  {Schoutens}}]{Zozulya:2008kb}%
  \BibitemOpen
  \bibfield  {author} {\bibinfo {author} {\bibfnamefont {O.~S.}\ \bibnamefont
  {Zozulya}}, \bibinfo {author} {\bibfnamefont {Masudul}\ \bibnamefont
  {Haque}}, \ and\ \bibinfo {author} {\bibfnamefont {K.}~\bibnamefont
  {Schoutens}},\ }\bibfield  {title} {\enquote {\bibinfo {title} {{Particle
  partitioning entanglement in itinerant many-particle systems}},}\ }\href
  {https://journals.aps.org/pra/abstract/10.1103/PhysRevA.78.042326} {\bibfield
   {journal} {\bibinfo  {journal} {Phys. Rev. A}\ }\textbf {\bibinfo {volume}
  {78}},\ \bibinfo {pages} {042326} (\bibinfo {year} {2008})}\BibitemShut
  {NoStop}%
\bibitem [{\citenamefont {Haque}\ \emph {et~al.}(2009)\citenamefont {Haque},
  \citenamefont {Zozulya},\ and\ \citenamefont {Schoutens}}]{Haque:2009df}%
  \BibitemOpen
  \bibfield  {author} {\bibinfo {author} {\bibfnamefont {Masudul}\ \bibnamefont
  {Haque}}, \bibinfo {author} {\bibfnamefont {O~S}\ \bibnamefont {Zozulya}}, \
  and\ \bibinfo {author} {\bibfnamefont {K}~\bibnamefont {Schoutens}},\
  }\bibfield  {title} {\enquote {\bibinfo {title} {{Entanglement between
  particle partitions in itinerant many-particle states}},}\ }\href
  {https://iopscience.iop.org/article/10.1088/1751-8113/42/50/504012}
  {\bibfield  {journal} {\bibinfo  {journal} {J. Phys. A: Math. Theor.}\
  }\textbf {\bibinfo {volume} {42}},\ \bibinfo {pages} {504012} (\bibinfo
  {year} {2009})}\BibitemShut {NoStop}%
\bibitem [{\citenamefont {Liu}\ and\ \citenamefont {Fan}(2010)}]{Liu:2010pe}%
  \BibitemOpen
  \bibfield  {author} {\bibinfo {author} {\bibfnamefont {Zhao}\ \bibnamefont
  {Liu}}\ and\ \bibinfo {author} {\bibfnamefont {Heng}\ \bibnamefont {Fan}},\
  }\bibfield  {title} {\enquote {\bibinfo {title} {{P}article entanglement in
  rotating gases},}\ }\href {\doibase 10.1103/physreva.81.062302} {\bibfield
  {journal} {\bibinfo  {journal} {Phys. Rev. A}\ }\textbf {\bibinfo {volume}
  {81}},\ \bibinfo {pages} {062302} (\bibinfo {year} {2010})}\BibitemShut
  {NoStop}%
\bibitem [{\citenamefont {Herdman}\ \emph
  {et~al.}(2014{\natexlab{a}})\citenamefont {Herdman}, \citenamefont {Roy},
  \citenamefont {Melko},\ and\ \citenamefont {Del~Maestro}}]{Herdman:2014jq}%
  \BibitemOpen
  \bibfield  {author} {\bibinfo {author} {\bibfnamefont {C.M.}\ \bibnamefont
  {Herdman}}, \bibinfo {author} {\bibfnamefont {P.N.}\ \bibnamefont {Roy}},
  \bibinfo {author} {\bibfnamefont {R.G.}\ \bibnamefont {Melko}}, \ and\
  \bibinfo {author} {\bibfnamefont {A.}~\bibnamefont {Del~Maestro}},\
  }\bibfield  {title} {\enquote {\bibinfo {title} {{Particle entanglement in
  continuum many-body systems via quantum Monte Carlo}},}\ }\href@noop {}
  {\bibfield  {journal} {\bibinfo  {journal} {Phys. Rev. B}\ }\textbf {\bibinfo
  {volume} {89}},\ \bibinfo {pages} {140501(R)} (\bibinfo {year}
  {2014}{\natexlab{a}})}\BibitemShut {NoStop}%
\bibitem [{\citenamefont {Herdman}\ \emph
  {et~al.}(2014{\natexlab{b}})\citenamefont {Herdman}, \citenamefont {Inglis},
  \citenamefont {Roy}, \citenamefont {Melko},\ and\ \citenamefont
  {Del~Maestro}}]{Herdman:2014ey}%
  \BibitemOpen
  \bibfield  {author} {\bibinfo {author} {\bibfnamefont {C.~M.}\ \bibnamefont
  {Herdman}}, \bibinfo {author} {\bibfnamefont {Stephen}\ \bibnamefont
  {Inglis}}, \bibinfo {author} {\bibfnamefont {P.~N.}\ \bibnamefont {Roy}},
  \bibinfo {author} {\bibfnamefont {R.~G.}\ \bibnamefont {Melko}}, \ and\
  \bibinfo {author} {\bibfnamefont {A.}~\bibnamefont {Del~Maestro}},\
  }\bibfield  {title} {\enquote {\bibinfo {title} {{Path-integral Monte Carlo
  method for R{\'e}nyi entanglement entropies}},}\ }\href@noop {} {\bibfield
  {journal} {\bibinfo  {journal} {Phys. Rev. E}\ }\textbf {\bibinfo {volume}
  {90}},\ \bibinfo {pages} {013308} (\bibinfo {year}
  {2014}{\natexlab{b}})}\BibitemShut {NoStop}%
\bibitem [{\citenamefont {Herdman}\ and\ \citenamefont
  {Del~Maestro}(2015)}]{Herdman:2015gx}%
  \BibitemOpen
  \bibfield  {author} {\bibinfo {author} {\bibfnamefont {C.~M.}\ \bibnamefont
  {Herdman}}\ and\ \bibinfo {author} {\bibfnamefont {A.}~\bibnamefont
  {Del~Maestro}},\ }\bibfield  {title} {\enquote {\bibinfo {title} {{Particle
  partition entanglement of bosonic Luttinger liquids}},}\ }\href@noop {}
  {\bibfield  {journal} {\bibinfo  {journal} {Phys. Rev. B}\ }\textbf {\bibinfo
  {volume} {91}},\ \bibinfo {pages} {184507} (\bibinfo {year}
  {2015})}\BibitemShut {NoStop}%
\bibitem [{\citenamefont {Rammelm{\"u}ller}\ \emph {et~al.}(2017)\citenamefont
  {Rammelm{\"u}ller}, \citenamefont {Porter}, \citenamefont {Braun},\ and\
  \citenamefont {Drut}}]{Rammelmuller:2017om}%
  \BibitemOpen
  \bibfield  {author} {\bibinfo {author} {\bibfnamefont {Lukas}\ \bibnamefont
  {Rammelm{\"u}ller}}, \bibinfo {author} {\bibfnamefont {William~J.}\
  \bibnamefont {Porter}}, \bibinfo {author} {\bibfnamefont {Jens}\ \bibnamefont
  {Braun}}, \ and\ \bibinfo {author} {\bibfnamefont {Joaqun~E.}\ \bibnamefont
  {Drut}},\ }\bibfield  {title} {\enquote {\bibinfo {title} {{E}volution from
  few- to many-body physics in one-dimensional {F}ermi systems: {O}ne- and
  two-body density matrices and particle-partition entanglement},}\ }\href
  {\doibase 10.1103/physreva.96.033635} {\bibfield  {journal} {\bibinfo
  {journal} {Phys. Rev. A}\ }\textbf {\bibinfo {volume} {96}},\ \bibinfo
  {pages} {033635} (\bibinfo {year} {2017})}\BibitemShut {NoStop}%
\bibitem [{\citenamefont {Barghathi}\ \emph {et~al.}(2017)\citenamefont
  {Barghathi}, \citenamefont {Casiano-Diaz},\ and\ \citenamefont
  {Del~Maestro}}]{Barghathi:2017fv}%
  \BibitemOpen
  \bibfield  {author} {\bibinfo {author} {\bibfnamefont {Hatem}\ \bibnamefont
  {Barghathi}}, \bibinfo {author} {\bibfnamefont {Emanuel}\ \bibnamefont
  {Casiano-Diaz}}, \ and\ \bibinfo {author} {\bibfnamefont {Adrian}\
  \bibnamefont {Del~Maestro}},\ }\bibfield  {title} {\enquote {\bibinfo {title}
  {{Particle partition entanglement of one dimensional spinless fermions}},}\
  }\href {\doibase 10.1088/1742-5468/aa819a} {\bibfield  {journal} {\bibinfo
  {journal} {J. Stat. Mech.: Theor. Exp.}\ }\textbf {\bibinfo {volume}
  {2017}},\ \bibinfo {pages} {083108} (\bibinfo {year} {2017})}\BibitemShut
  {NoStop}%
\bibitem [{\citenamefont {Hopjan}\ \emph {et~al.}(2021)\citenamefont {Hopjan},
  \citenamefont {Heidrich-Meisner},\ and\ \citenamefont
  {Alba}}]{Hopjan:2020sp}%
  \BibitemOpen
  \bibfield  {author} {\bibinfo {author} {\bibfnamefont {Miroslav}\
  \bibnamefont {Hopjan}}, \bibinfo {author} {\bibfnamefont {Fabian}\
  \bibnamefont {Heidrich-Meisner}}, \ and\ \bibinfo {author} {\bibfnamefont
  {Vincenzo}\ \bibnamefont {Alba}},\ }\bibfield  {title} {\enquote {\bibinfo
  {title} {Scaling properties of a spatial one-particle density-matrix entropy
  in many-body localized systems},}\ }\href {\doibase
  10.1103/PhysRevB.104.035129} {\bibfield  {journal} {\bibinfo  {journal}
  {Phys. Rev. B}\ }\textbf {\bibinfo {volume} {104}},\ \bibinfo {pages}
  {035129} (\bibinfo {year} {2021})}\BibitemShut {NoStop}%
\bibitem [{\citenamefont {L{\"o}wdin}(1955)}]{Lowdin:1955uo}%
  \BibitemOpen
  \bibfield  {author} {\bibinfo {author} {\bibfnamefont {Per-Olov}\
  \bibnamefont {L{\"o}wdin}},\ }\bibfield  {title} {\enquote {\bibinfo {title}
  {{Q}uantum {T}heory of {M}any-{P}article {S}ystems. {I}. {P}hysical
  {I}nterpretations by {M}eans of {D}ensity {M}atrices, {N}atural
  {S}pin-{O}rbitals, and {C}onvergence {P}roblems in the {M}ethod of
  {C}onfigurational {I}nteraction},}\ }\href {\doibase 10.1103/physrev.97.1474}
  {\bibfield  {journal} {\bibinfo  {journal} {Phys. Rev.}\ }\textbf {\bibinfo
  {volume} {97}},\ \bibinfo {pages} {1474} (\bibinfo {year}
  {1955})}\BibitemShut {NoStop}%
\bibitem [{\citenamefont {Coleman}(1963)}]{Coleman:1963lt}%
  \BibitemOpen
  \bibfield  {author} {\bibinfo {author} {\bibfnamefont {A.~J.}\ \bibnamefont
  {Coleman}},\ }\bibfield  {title} {\enquote {\bibinfo {title} {{S}tructure of
  {F}ermion {D}ensity {M}atrices},}\ }\href {\doibase
  10.1103/revmodphys.35.668} {\bibfield  {journal} {\bibinfo  {journal} {Rev.
  Mod. Phys.}\ }\textbf {\bibinfo {volume} {35}},\ \bibinfo {pages} {668}
  (\bibinfo {year} {1963})}\BibitemShut {NoStop}%
\bibitem [{\citenamefont {Mahan}(2000)}]{Mahan:2000mp}%
  \BibitemOpen
  \bibfield  {author} {\bibinfo {author} {\bibfnamefont {G.D.}\ \bibnamefont
  {Mahan}},\ }\href@noop {} {\emph {\bibinfo {title} {Many-Particle Physics}}}\
  (\bibinfo  {publisher} {Kluwer Academic/Plenum Publishers},\ \bibinfo
  {address} {New York},\ \bibinfo {year} {2000})\BibitemShut {NoStop}%
\bibitem [{\citenamefont {Hastings}(2007)}]{Hastings:2007bu}%
  \BibitemOpen
  \bibfield  {author} {\bibinfo {author} {\bibfnamefont {M.~B.}\ \bibnamefont
  {Hastings}},\ }\bibfield  {title} {\enquote {\bibinfo {title} {{An area law
  for one-dimensional quantum systems}},}\ }\href
  {https://iopscience.iop.org/article/10.1088/1742-5468/2007/08/P08024}
  {\bibfield  {journal} {\bibinfo  {journal} {J. Stat. Mech.: Theor. Exp.}\
  }\textbf {\bibinfo {volume} {2007}},\ \bibinfo {pages} {P08024} (\bibinfo
  {year} {2007})}\BibitemShut {NoStop}%
\bibitem [{\citenamefont {Eisert}\ \emph {et~al.}(2010)\citenamefont {Eisert},
  \citenamefont {Cramer},\ and\ \citenamefont {Plenio}}]{Eisert:2010hq}%
  \BibitemOpen
  \bibfield  {author} {\bibinfo {author} {\bibfnamefont {J.}~\bibnamefont
  {Eisert}}, \bibinfo {author} {\bibfnamefont {M.}~\bibnamefont {Cramer}}, \
  and\ \bibinfo {author} {\bibfnamefont {M.~B.}\ \bibnamefont {Plenio}},\
  }\bibfield  {title} {\enquote {\bibinfo {title} {{Colloquium: Area laws for
  the entanglement entropy}},}\ }\href {\doibase 10.1103/RevModPhys.82.277}
  {\bibfield  {journal} {\bibinfo  {journal} {Rev. Mod. Phys.}\ }\textbf
  {\bibinfo {volume} {82}},\ \bibinfo {pages} {277} (\bibinfo {year}
  {2010})}\BibitemShut {NoStop}%
\bibitem [{\citenamefont {Pastori}\ \emph {et~al.}(2019)\citenamefont
  {Pastori}, \citenamefont {Heyl},\ and\ \citenamefont
  {Budich}}]{Pastori:2019ds}%
  \BibitemOpen
  \bibfield  {author} {\bibinfo {author} {\bibfnamefont {Lorenzo}\ \bibnamefont
  {Pastori}}, \bibinfo {author} {\bibfnamefont {Markus}\ \bibnamefont {Heyl}},
  \ and\ \bibinfo {author} {\bibfnamefont {Jan~Carl}\ \bibnamefont {Budich}},\
  }\bibfield  {title} {\enquote {\bibinfo {title} {Disentangling sources of
  quantum entanglement in quench dynamics},}\ }\href
  {https://link.aps.org/doi/10.1103/PhysRevResearch.1.012007} {\bibfield
  {journal} {\bibinfo  {journal} {Phys. Rev. Research}\ }\textbf {\bibinfo
  {volume} {1}},\ \bibinfo {pages} {012007} (\bibinfo {year}
  {2019})}\BibitemShut {NoStop}%
\bibitem [{\citenamefont {Lundgren}\ \emph {et~al.}(2019)\citenamefont
  {Lundgren}, \citenamefont {Liu}, \citenamefont {Laurell},\ and\ \citenamefont
  {Fiete}}]{Lundgren:2019aa}%
  \BibitemOpen
  \bibfield  {author} {\bibinfo {author} {\bibfnamefont {Rex}\ \bibnamefont
  {Lundgren}}, \bibinfo {author} {\bibfnamefont {Fangli}\ \bibnamefont {Liu}},
  \bibinfo {author} {\bibfnamefont {Pontus}\ \bibnamefont {Laurell}}, \ and\
  \bibinfo {author} {\bibfnamefont {Gregory~A.}\ \bibnamefont {Fiete}},\
  }\bibfield  {title} {\enquote {\bibinfo {title} {{M}omentum-space
  entanglement after a quench in one-dimensional disordered fermionic
  systems},}\ }\href {\doibase 10.1103/physrevb.100.241108} {\bibfield
  {journal} {\bibinfo  {journal} {Phys. Rev. B}\ }\textbf {\bibinfo {volume}
  {100}},\ \bibinfo {pages} {241108} (\bibinfo {year} {2019})}\BibitemShut
  {NoStop}%
\bibitem [{\citenamefont {Cloizeaux}(1966)}]{DesCloizeaux:1966}%
  \BibitemOpen
  \bibfield  {author} {\bibinfo {author} {\bibfnamefont {J.~Des}\ \bibnamefont
  {Cloizeaux}},\ }\bibfield  {title} {\enquote {\bibinfo {title} {{A Soluble
  Fermi-Gas Model. Validity of Transformations of the Bogoliubov Type}},}\
  }\href {\doibase 10.1063/1.1704899} {\bibfield  {journal} {\bibinfo
  {journal} {J. Math. Phys.}\ }\textbf {\bibinfo {volume} {7}},\ \bibinfo
  {pages} {2136} (\bibinfo {year} {1966})}\BibitemShut {NoStop}%
\bibitem [{\citenamefont {Yang}\ and\ \citenamefont
  {Yang}(1966)}]{Yang:1966yc}%
  \BibitemOpen
  \bibfield  {author} {\bibinfo {author} {\bibfnamefont {C.~N.}\ \bibnamefont
  {Yang}}\ and\ \bibinfo {author} {\bibfnamefont {C.~P.}\ \bibnamefont
  {Yang}},\ }\bibfield  {title} {\enquote {\bibinfo {title}
  {{O}ne-{D}imensional {C}hain of {A}nisotropic {S}pin-{S}pin {I}nteractions.
  {I}. {P}roof of {B}ethe's {H}ypothesis for {G}round {S}tate in a {F}inite
  {S}ystem},}\ }\href {\doibase 10.1103/physrev.150.321} {\bibfield  {journal}
  {\bibinfo  {journal} {Phys. Rev.}\ }\textbf {\bibinfo {volume} {150}},\
  \bibinfo {pages} {321} (\bibinfo {year} {1966})}\BibitemShut {NoStop}%
\bibitem [{rep(2020)}]{repo}%
  \BibitemOpen
  \href {\doibase doi:10.5281/zenodo.3634078} {} (\bibinfo {year} {2020}),\
  \bibinfo {note} {{All code, scripts and data used in this work are included
  in a GitHub repository:
  \url{https://github.com/DelMaestroGroup/papers-code-EntanglementQuantumQuench},
  doi:10.5281/zenodo.3634078}}\BibitemShut {NoStop}%
\bibitem [{\citenamefont {Fioretto}\ and\ \citenamefont
  {Mussardo}(2010)}]{Fioretto:2010qc}%
  \BibitemOpen
  \bibfield  {author} {\bibinfo {author} {\bibfnamefont {Davide}\ \bibnamefont
  {Fioretto}}\ and\ \bibinfo {author} {\bibfnamefont {Giuseppe}\ \bibnamefont
  {Mussardo}},\ }\bibfield  {title} {\enquote {\bibinfo {title} {Quantum
  quenches in integrable field theories},}\ }\href {\doibase
  10.1088/1367-2630/12/5/055015} {\bibfield  {journal} {\bibinfo  {journal}
  {New. J. Phys.}\ }\textbf {\bibinfo {volume} {12}},\ \bibinfo {pages}
  {055015} (\bibinfo {year} {2010})}\BibitemShut {NoStop}%
\bibitem [{\citenamefont {St{\'{e}}phan}\ and\ \citenamefont
  {Dubail}(2011)}]{Stephan:2011lq}%
  \BibitemOpen
  \bibfield  {author} {\bibinfo {author} {\bibfnamefont {Jean-Marie}\
  \bibnamefont {St{\'{e}}phan}}\ and\ \bibinfo {author} {\bibfnamefont
  {J{\'{e}}r{\^{o}}me}\ \bibnamefont {Dubail}},\ }\bibfield  {title} {\enquote
  {\bibinfo {title} {Local quantum quenches in critical one-dimensional
  systems: entanglement, the loschmidt echo, and light-cone effects},}\ }\href
  {\doibase 10.1088/1742-5468/2011/08/p08019} {\bibfield  {journal} {\bibinfo
  {journal} {J. Stat. Mech: Theor. Exp.}\ }\textbf {\bibinfo {volume} {2011}},\
  \bibinfo {pages} {P08019} (\bibinfo {year} {2011})}\BibitemShut {NoStop}%
\bibitem [{\citenamefont {L{\"a}uchli}\ and\ \citenamefont
  {Kollath}(2008)}]{Lauchli:2008wb}%
  \BibitemOpen
  \bibfield  {author} {\bibinfo {author} {\bibfnamefont {Andreas~M.}\
  \bibnamefont {L{\"a}uchli}}\ and\ \bibinfo {author} {\bibfnamefont {Corinna}\
  \bibnamefont {Kollath}},\ }\bibfield  {title} {\enquote {\bibinfo {title}
  {{S}preading of correlations and entanglement after a quench in the
  one-dimensional {B}ose-{H}ubbard model},}\ }\href {\doibase
  10.1088/1742-5468/2008/05/p05018} {\bibfield  {journal} {\bibinfo  {journal}
  {J. Stat. Mech.: Theor. Exp.}\ }\textbf {\bibinfo {volume} {2008}},\ \bibinfo
  {pages} {P05018} (\bibinfo {year} {2008})}\BibitemShut {NoStop}%
\bibitem [{\citenamefont {Calabrese}\ and\ \citenamefont
  {Cardy}(2016)}]{Calabrese:2016sn}%
  \BibitemOpen
  \bibfield  {author} {\bibinfo {author} {\bibfnamefont {Pasquale}\
  \bibnamefont {Calabrese}}\ and\ \bibinfo {author} {\bibfnamefont {John}\
  \bibnamefont {Cardy}},\ }\bibfield  {title} {\enquote {\bibinfo {title}
  {{Q}uantum quenches in 1+1 dimensional conformal field theories},}\ }\href
  {\doibase 10.1088/1742-5468/2016/06/064003} {\bibfield  {journal} {\bibinfo
  {journal} {J. Stat. Mech.: Theor. Exp.}\ }\textbf {\bibinfo {volume}
  {2016}},\ \bibinfo {pages} {064003} (\bibinfo {year} {2016})}\BibitemShut
  {NoStop}%
\bibitem [{\citenamefont {Garrison}\ and\ \citenamefont
  {Grover}(2018)}]{Garrison:2018kv}%
  \BibitemOpen
  \bibfield  {author} {\bibinfo {author} {\bibfnamefont {James~R.}\
  \bibnamefont {Garrison}}\ and\ \bibinfo {author} {\bibfnamefont {Tarun}\
  \bibnamefont {Grover}},\ }\bibfield  {title} {\enquote {\bibinfo {title}
  {Does a single eigenstate encode the full hamiltonian?}}\ }\href
  {https://link.aps.org/doi/10.1103/PhysRevX.8.021026} {\bibfield  {journal}
  {\bibinfo  {journal} {Phys. Rev. X}\ }\textbf {\bibinfo {volume} {8}},\
  \bibinfo {pages} {021026} (\bibinfo {year} {2018})}\BibitemShut {NoStop}%
\bibitem [{\citenamefont {Jin}\ and\ \citenamefont
  {Korepin}(2004)}]{Jin:2004uy}%
  \BibitemOpen
  \bibfield  {author} {\bibinfo {author} {\bibfnamefont {B.~Q.}\ \bibnamefont
  {Jin}}\ and\ \bibinfo {author} {\bibfnamefont {V.~E.}\ \bibnamefont
  {Korepin}},\ }\bibfield  {title} {\enquote {\bibinfo {title} {{Q}uantum
  {S}pin {C}hain, {T}oeplitz {D}eterminants and the
  {F}isher{\textendash}{H}artwig {C}onjecture},}\ }\href {\doibase
  10.1023/b:joss.0000037230.37166.42} {\bibfield  {journal} {\bibinfo
  {journal} {J. Stat. Phys.}\ }\textbf {\bibinfo {volume} {116}},\ \bibinfo
  {pages} {79} (\bibinfo {year} {2004})}\BibitemShut {NoStop}%
\bibitem [{\citenamefont {Ghirardi}\ and\ \citenamefont
  {Marinatto}(2005)}]{Ghirardi:2005md}%
  \BibitemOpen
  \bibfield  {author} {\bibinfo {author} {\bibfnamefont {G.~C.}\ \bibnamefont
  {Ghirardi}}\ and\ \bibinfo {author} {\bibfnamefont {L.}~\bibnamefont
  {Marinatto}},\ }\bibfield  {title} {\enquote {\bibinfo {title} {{I}dentical
  {P}articles and {E}ntanglement},}\ }\href {\doibase 10.1134/1.2055932}
  {\bibfield  {journal} {\bibinfo  {journal} {Opt. Spectrosc.}\ }\textbf
  {\bibinfo {volume} {99}},\ \bibinfo {pages} {386} (\bibinfo {year}
  {2005})}\BibitemShut {NoStop}%
\bibitem [{\citenamefont {Carlen}\ \emph {et~al.}(2016)\citenamefont {Carlen},
  \citenamefont {Lieb},\ and\ \citenamefont {Reuvers}}]{Carlen:2016es}%
  \BibitemOpen
  \bibfield  {author} {\bibinfo {author} {\bibfnamefont {Eric~A.}\ \bibnamefont
  {Carlen}}, \bibinfo {author} {\bibfnamefont {Elliott~H.}\ \bibnamefont
  {Lieb}}, \ and\ \bibinfo {author} {\bibfnamefont {Robin}\ \bibnamefont
  {Reuvers}},\ }\bibfield  {title} {\enquote {\bibinfo {title} {{E}ntropy and
  {E}ntanglement {B}ounds for {R}educed {D}ensity {M}atrices of {F}ermionic
  {S}tates},}\ }\href {\doibase 10.1007/s00220-016-2651-6} {\bibfield
  {journal} {\bibinfo  {journal} {Commun. Math. Phys.}\ }\textbf {\bibinfo
  {volume} {344}},\ \bibinfo {pages} {655} (\bibinfo {year}
  {2016})}\BibitemShut {NoStop}%
\bibitem [{\citenamefont {Lemm}(2017)}]{Lemm:2017ef}%
  \BibitemOpen
  \bibfield  {author} {\bibinfo {author} {\bibfnamefont {Marius}\ \bibnamefont
  {Lemm}},\ }\bibfield  {title} {\enquote {\bibinfo {title} {On the entropy of
  fermionic reduced density matrices},}\ }\href
  {https://arxiv.org/abs/1702.02360} {\  (\bibinfo {year} {2017})},\ \Eprint
  {http://arxiv.org/abs/1702.02360} {arXiv:1702.02360 [quant-ph]} \BibitemShut
  {NoStop}%
\bibitem [{\citenamefont {Herdman}\ \emph {et~al.}(2016)\citenamefont
  {Herdman}, \citenamefont {Roy}, \citenamefont {Melko},\ and\ \citenamefont
  {{Del Maestro}}}]{Herdman:2016ep}%
  \BibitemOpen
  \bibfield  {author} {\bibinfo {author} {\bibfnamefont {C.~M.}\ \bibnamefont
  {Herdman}}, \bibinfo {author} {\bibfnamefont {P.~N.}\ \bibnamefont {Roy}},
  \bibinfo {author} {\bibfnamefont {R.~G.}\ \bibnamefont {Melko}}, \ and\
  \bibinfo {author} {\bibfnamefont {A.}~\bibnamefont {{Del Maestro}}},\
  }\bibfield  {title} {\enquote {\bibinfo {title} {{S}patial entanglement
  entropy in the ground state of the {L}ieb-{L}iniger model},}\ }\href
  {\doibase 10.1103/physrevb.94.064524} {\bibfield  {journal} {\bibinfo
  {journal} {Phys. Rev. B}\ }\textbf {\bibinfo {volume} {94}},\ \bibinfo
  {pages} {064524} (\bibinfo {year} {2016})}\BibitemShut {NoStop}%
\bibitem [{\citenamefont {Piroli}\ \emph {et~al.}(2017)\citenamefont {Piroli},
  \citenamefont {Vernier}, \citenamefont {Calabrese},\ and\ \citenamefont
  {Rigol}}]{Piroli:2017un}%
  \BibitemOpen
  \bibfield  {author} {\bibinfo {author} {\bibfnamefont {Lorenzo}\ \bibnamefont
  {Piroli}}, \bibinfo {author} {\bibfnamefont {Eric}\ \bibnamefont {Vernier}},
  \bibinfo {author} {\bibfnamefont {Pasquale}\ \bibnamefont {Calabrese}}, \
  and\ \bibinfo {author} {\bibfnamefont {Marcos}\ \bibnamefont {Rigol}},\
  }\bibfield  {title} {\enquote {\bibinfo {title} {Correlations and diagonal
  entropy after quantum quenches in {XXZ} chains},}\ }\href
  {https://link.aps.org/doi/10.1103/PhysRevB.95.054308} {\bibfield  {journal}
  {\bibinfo  {journal} {Phys. Rev. B}\ }\textbf {\bibinfo {volume} {95}},\
  \bibinfo {pages} {054308} (\bibinfo {year} {2017})}\BibitemShut {NoStop}%
\bibitem [{\citenamefont {Mishra}\ \emph {et~al.}(2011)\citenamefont {Mishra},
  \citenamefont {Carrasquilla},\ and\ \citenamefont {Rigol}}]{Mishra:2011gd}%
  \BibitemOpen
  \bibfield  {author} {\bibinfo {author} {\bibfnamefont {Tapan}\ \bibnamefont
  {Mishra}}, \bibinfo {author} {\bibfnamefont {Juan}\ \bibnamefont
  {Carrasquilla}}, \ and\ \bibinfo {author} {\bibfnamefont {Marcos}\
  \bibnamefont {Rigol}},\ }\bibfield  {title} {\enquote {\bibinfo {title}
  {{P}hase diagram of the half-filled one-dimensional $t-{V}-{V}'$ model},}\
  }\href {\doibase 10.1103/physrevb.84.115135} {\bibfield  {journal} {\bibinfo
  {journal} {Phys. Rev. B}\ }\textbf {\bibinfo {volume} {84}},\ \bibinfo
  {pages} {115135} (\bibinfo {year} {2011})}\BibitemShut {NoStop}%
\bibitem [{\citenamefont {White}(1992)}]{White:1992rs}%
  \BibitemOpen
  \bibfield  {author} {\bibinfo {author} {\bibfnamefont {Steven~R.}\
  \bibnamefont {White}},\ }\bibfield  {title} {\enquote {\bibinfo {title}
  {{D}ensity matrix formulation for quantum renormalization groups},}\ }\href
  {\doibase 10.1103/physrevlett.69.2863} {\bibfield  {journal} {\bibinfo
  {journal} {Phys. Rev. Lett.}\ }\textbf {\bibinfo {volume} {69}},\ \bibinfo
  {pages} {2863} (\bibinfo {year} {1992})}\BibitemShut {NoStop}%
\bibitem [{\citenamefont {Schollw{\"o}ck}(2011)}]{Schollwoeck:2011jf}%
  \BibitemOpen
  \bibfield  {author} {\bibinfo {author} {\bibfnamefont {Ulrich}\ \bibnamefont
  {Schollw{\"o}ck}},\ }\bibfield  {title} {\enquote {\bibinfo {title} {{T}he
  density-matrix renormalization group in the age of matrix product states},}\
  }\href {\doibase 10.1016/j.aop.2010.09.012} {\bibfield  {journal} {\bibinfo
  {journal} {Ann. Phys.}\ }\textbf {\bibinfo {volume} {326}},\ \bibinfo {pages}
  {96} (\bibinfo {year} {2011})}\BibitemShut {NoStop}%
\bibitem [{\citenamefont {Kurashige}\ \emph {et~al.}(2014)\citenamefont
  {Kurashige}, \citenamefont {Chalupsk{\'{y}}}, \citenamefont {Lan},\ and\
  \citenamefont {Yanai}}]{Kurashige:2014ca}%
  \BibitemOpen
  \bibfield  {author} {\bibinfo {author} {\bibfnamefont {Yuki}\ \bibnamefont
  {Kurashige}}, \bibinfo {author} {\bibfnamefont {Jakub}\ \bibnamefont
  {Chalupsk{\'{y}}}}, \bibinfo {author} {\bibfnamefont {Tran~Nguyen}\
  \bibnamefont {Lan}}, \ and\ \bibinfo {author} {\bibfnamefont {Takeshi}\
  \bibnamefont {Yanai}},\ }\bibfield  {title} {\enquote {\bibinfo {title}
  {Complete active space second-order perturbation theory with cumulant
  approximation for extended active-space wavefunction from density matrix
  renormalization group},}\ }\href {\doibase 10.1063/1.4900878} {\bibfield
  {journal} {\bibinfo  {journal} {J. Chem. Phys.}\ }\textbf {\bibinfo {volume}
  {141}},\ \bibinfo {pages} {174111} (\bibinfo {year} {2014})}\BibitemShut
  {NoStop}%
\bibitem [{\citenamefont {Fetter}\ and\ \citenamefont {Walecka}(2003)}]{FeWa}%
  \BibitemOpen
  \bibfield  {author} {\bibinfo {author} {\bibfnamefont {A.}~\bibnamefont
  {Fetter}}\ and\ \bibinfo {author} {\bibfnamefont {J.~D.}\ \bibnamefont
  {Walecka}},\ }\href@noop {} {\emph {\bibinfo {title} {Quantum Theory of
  Many-Particle Systems}}}\ (\bibinfo  {publisher} {Dover Publications},\
  \bibinfo {address} {Mineola, NY},\ \bibinfo {year} {2003})\BibitemShut
  {NoStop}%
\bibitem [{\citenamefont {Green}(1952)}]{Green52}%
  \BibitemOpen
  \bibfield  {author} {\bibinfo {author} {\bibfnamefont {H.~S.}\ \bibnamefont
  {Green}},\ }\href@noop {} {\emph {\bibinfo {title} {The Molecular Theory of
  Fluids}}}\ (\bibinfo  {publisher} {Norh-Holland},\ \bibinfo {address}
  {Amsterdam},\ \bibinfo {year} {1952})\BibitemShut {NoStop}%
\bibitem [{\citenamefont {Kirkwood}\ and\ \citenamefont
  {Boggs}(1942)}]{Kirkwood42}%
  \BibitemOpen
  \bibfield  {author} {\bibinfo {author} {\bibfnamefont {John~G.}\ \bibnamefont
  {Kirkwood}}\ and\ \bibinfo {author} {\bibfnamefont {Elizabeth~Monroe}\
  \bibnamefont {Boggs}},\ }\bibfield  {title} {\enquote {\bibinfo {title}
  {{T}he {R}adial {D}istribution {F}unction in {L}iquids},}\ }\href {\doibase
  10.1063/1.1723737} {\bibfield  {journal} {\bibinfo  {journal} {J. Chem.
  Phys.}\ }\textbf {\bibinfo {volume} {10}},\ \bibinfo {pages} {394} (\bibinfo
  {year} {1942})}\BibitemShut {NoStop}%
\bibitem [{\citenamefont {Wallace}(1987)}]{Wallace87}%
  \BibitemOpen
  \bibfield  {author} {\bibinfo {author} {\bibfnamefont {Duane~C.}\
  \bibnamefont {Wallace}},\ }\bibfield  {title} {\enquote {\bibinfo {title}
  {{O}n the role of density fluctuations in the entropy of a fluid},}\ }\href
  {\doibase 10.1063/1.453158} {\bibfield  {journal} {\bibinfo  {journal} {J.
  Chem. Phys.}\ }\textbf {\bibinfo {volume} {87}},\ \bibinfo {pages} {2282}
  (\bibinfo {year} {1987})}\BibitemShut {NoStop}%
\bibitem [{\citenamefont {Evans}(1989)}]{Evans89}%
  \BibitemOpen
  \bibfield  {author} {\bibinfo {author} {\bibfnamefont {Denis~J.}\
  \bibnamefont {Evans}},\ }\bibfield  {title} {\enquote {\bibinfo {title} {{O}n
  the entropy of nonequilibrium states},}\ }\href {\doibase 10.1007/bf01022830}
  {\bibfield  {journal} {\bibinfo  {journal} {J. Stat. Phys.}\ }\textbf
  {\bibinfo {volume} {57}},\ \bibinfo {pages} {745} (\bibinfo {year}
  {1989})}\BibitemShut {NoStop}%
\end{thebibliography}%
\end{document}